\documentclass[preprint, hidelinks, 10pt]{elsarticle}

\usepackage{amsmath}
\usepackage{tcolorbox}
\usepackage{upgreek}

\usepackage{graphicx}
\usepackage[version=4]{mhchem}
\usepackage{amssymb}

\usepackage{color}

\let\mpar=\marginpar
\renewcommand\marginpar[1]{\mpar{\raggedright \scriptsize #1}}

\usepackage[labelfont=bf,labelsep=period,justification=raggedright]{caption}

\makeatletter
\renewcommand{\@biblabel}[1]{\quad#1.}
\makeatother
\date{}

\usepackage{graphicx}
\usepackage{dcolumn}%
\usepackage{bm}%
\usepackage{hyperref}
\usepackage{float}

\def\be{\begin{equation}}   \def\ee{\end{equation}}

\usepackage{xr}
\begin{document}

\begin{center}
{\Large
\textbf{Wildebeest Herds on Rolling Hills:\\ Flocking on Arbitrary Curved Surfaces}
}
\bigskip

\bf{Christina L. Hueschen}$^{1,\ast}$, \bf{Alexander R. Dunn}$^{1}$,
\bf{Rob Phillips}$^{2,3,\ast}$
\\
\smallskip

{1} Department of Chemical Engineering, Stanford University, Palo Alto, California, 94305
\\
{2} Department of Physics, California Institute of Technology, Pasadena, California, 91125
\\
{3} Division of Biology and Biological Engineering, California Institute of Technology, Pasadena, California, 91125\\
$\ast$ Correspondence: chueschen@gmail.com, phillips@pboc.caltech.edu
\end{center}

\tableofcontents

\newpage
\noindent {\bf Abstract}
\addcontentsline{toc}{section}{Abstract}

The collective behavior of active agents, whether herds of wildebeest or microscopic actin filaments propelled by molecular motors, is an exciting frontier in biological and soft matter physics. 
Almost three decades ago, Toner and Tu developed a continuum theory of the collective action of flocks, or herds, that helped launch the modern field of active matter. 
One challenge faced when applying continuum active matter theories to living phenomena is the complex geometric structure of biological environments. Both macroscopic and microscopic herds move on asymmetric curved surfaces, like undulating grass plains or the surface layers of cells or embryos, which can render problems analytically intractable. In this work, 
we present a formulation of the Toner-Tu flocking theory that uses the finite element method to solve the governing equations on arbitrary curved surfaces. 
First, we test the developed formalism and its numerical implementation in channel flow with scattering obstacles and on cylindrical and spherical surfaces, comparing our results to analytical solutions. We then progress to surfaces with arbitrary curvature, moving beyond previously accessible problems to explore herding behavior on a variety of landscapes. This approach allows the investigation of transients and dynamic solutions not revealed by analytic methods. It also enables versatile incorporation of new geometries and boundary conditions and efficient sweeps of parameter space. Looking forward, the work presented here lays the groundwork for a dialogue between Toner-Tu theory and data on collective motion in biologically-relevant geometries,
from drone footage of migrating animal herds to movies of microscopic cytoskeletal flows within cells.

\section{Introduction}
The beautiful collective motions of flocking or herding animals have mesmerized the human mind for millennia \cite{Pliny1949} and inspired the modern field of study of active matter~\cite{Vicsek1995,Toner1995,Toner1998}.  
In recent decades, active matter theories have been used to describe collective motion in living systems across nearly a billion-fold difference in scales \cite{Marchetti2013}, from starling flocks wheeling over Rome at 10 m/s~\cite{Ballerini2008}, to 50 $\upmu$m/s flows of cytoplasm in cm-size internodal cells of the algae {\it Chara}~\cite{Woodhouse2013}, to 0.1 $\upmu$m/s flows of actin filaments and myosin molecules in 50 $\upmu$m developing {\it C. elegans} worm embryos~\cite{Mayer2010}. In these macroscopic and microscopic contexts, beautiful self-organization phenomena often take place on surfaces. The collective motion of flocking sheep or kilometer-wide migrating wildebeest herds are influenced by the hills, valleys, or canyons of the landscape on which they travel. Similarly, the flows of actin and myosin molecules mentioned above occur in thin layers at the surface of the embryo or cell, and are thus constrained by its surface topology and shape. 
While modern techniques such as dynamical drone imaging \cite{Cavagna2014} and high-resolution fluorescence microscopy \cite{Munster2019, Pimpale2020} enable experimental measurement of these animal, cell, or molecule kinematics, a dialogue between measurement and theory requires predictions that incorporate the complex shapes of the real-world surfaces on which they move.

In this work, we present a general curved-surface formulation and numerical implementation of a minimalistic continuum theory of flocking or herding active matter, with the hope that it will prove useful to others interested in exploring continuum theory predictions on complex geometries.
For active matter systems such as bacterial swarms \cite{Wioland2016}, active colloidal fluids \cite{Bricard2013}, self-propelled rods \cite{Kumar2014}, or purified cytoskeletal networks \cite{Sanchez2012}, studying the contribution of engineered confinement geometries to emergent patterns has proven a fruitful path \cite{Morin2017, Souslov2017,  Wu2017, Shankar2020, Zhang2020a}. Understanding the contribution of \textit{biological} geometries to pattern formation is an exciting direction of growth \cite{Behrndt2012, Streichan2018, Maroudas-Sacks2021, Cicconofri2020, Pearce2020, Hoffmann2022}.
Indeed, our own interest in solving active matter theories on arbitrary curved surfaces was initially inspired by a need to predict emergent actin polarity patterns at the surface of single-celled {\it Toxoplasma gondii} parasites, whose gliding motility is driven and directed by this surface actin layer~\cite{Hueschen2022c}.  These cells have  a beautiful but complex shape which lacks the symmetries that license traditional analytic approaches.
We focus on a classic continuum active matter model originally developed by John Toner and Yuhai Tu~\cite{Toner1995,Toner1998,Toner2005}, inspired by the work of T\'{a}mas Vicsek {\it et al.} \cite{Vicsek1995}, to describe the collective behavior of flocking or herding animals.  The Toner-Tu theory helped launch the modern field of active matter \cite{Bowick2022} and can
describe collections of dry, polar, self-propelled agents at any length scale.
We develop a general curved surface framework for the Toner-Tu theory and implement it in the finite element setting, enabling its convenient use on complex surfaces. While flocking theory - and in particular the herding of wildebeest - serves as our example in this study, the general curved surface formulation and finite element method (FEM) approach presented here may prove useful for any continuum active matter theory.
There is a rich and beautiful literature associated with formulating
field equations on arbitrary curved surfaces~\cite{Scriven1960, Fily2016, Voigt2017,Sahu2017,Jankuhn2018, Mietke2019,TorresSanchez2019}.
It has been said of these formulations, ``The complexity of the equations may explain why they are so often written but never solved for arbitrary surfaces"~\cite{Jankuhn2018}.  Because of 
our interest in collective motions on biological surfaces, we could not afford 
to avoid solving these equations on such surfaces.
We use the finite element method, which can be flexibly adjusted to an arbitrary choice of geometry and permits an exploration of active, self-organized solutions predicted by the Toner-Tu theory.  With our general surface formulation and finite element implementation in hand, we test our framework and approach on 
cylindrical and spherical geometries for which analytic solutions exist as well.  Satisfied by the agreement between our numerical and analytic results, we explore flocking phenomena on a broad collection of surfaces.

The remainder of the paper is organized as follows.
In Section~\ref{Section:FlockingTheory}, we show how the flocking theory of Toner and Tu can
be recast in a general surface form to describe flocking motions on arbitrary curved surfaces.  
In Section~\ref{Section:FiniteElement},
we construct a finite element surface formulation that respects the
low symmetry of realistic curved geometries while permitting numerical analysis of
the dynamics.  In Section~\ref{Section:ParametersSection}, we turn to the great wildebeest herds for inspiration and perform simple parameter estimates for Toner-Tu wildebeest herding. We then explore
the dimensionless ratios that appear when recasting the Toner-Tu equations in dimensionless form. In
Section~\ref{Section:AnalyticSolutions}, with the full surface formulation and its numerical implementation in hand, we explore scaling relationships and changes of dynamical state that arise for herds in channels with scattering obstacles, and we use cylindrical and spherical surfaces to compare our finite element method results to corresponding analytic solutions. Our approach also allows the exploration of transients and dynamic solutions not revealed by analytic methods.  Finally, in Section~\ref{section:BiologicalGeometries},
we playfully use the curved-space formalism and its finite element implementation for case studies of wildebeest herding on landscapes of rolling hills, and we note that this work serves as a theoretical foundation to solve flocking theories on real-world curved surfaces of biological interest.

\section{Flocking Theory for Arbitrary Curved Surfaces}
\label{Section:FlockingTheory}

\subsection{A Minimal Toner-Tu Theory in the Plane}

\begin{figure}
\centering{\includegraphics[width=6.0truein]{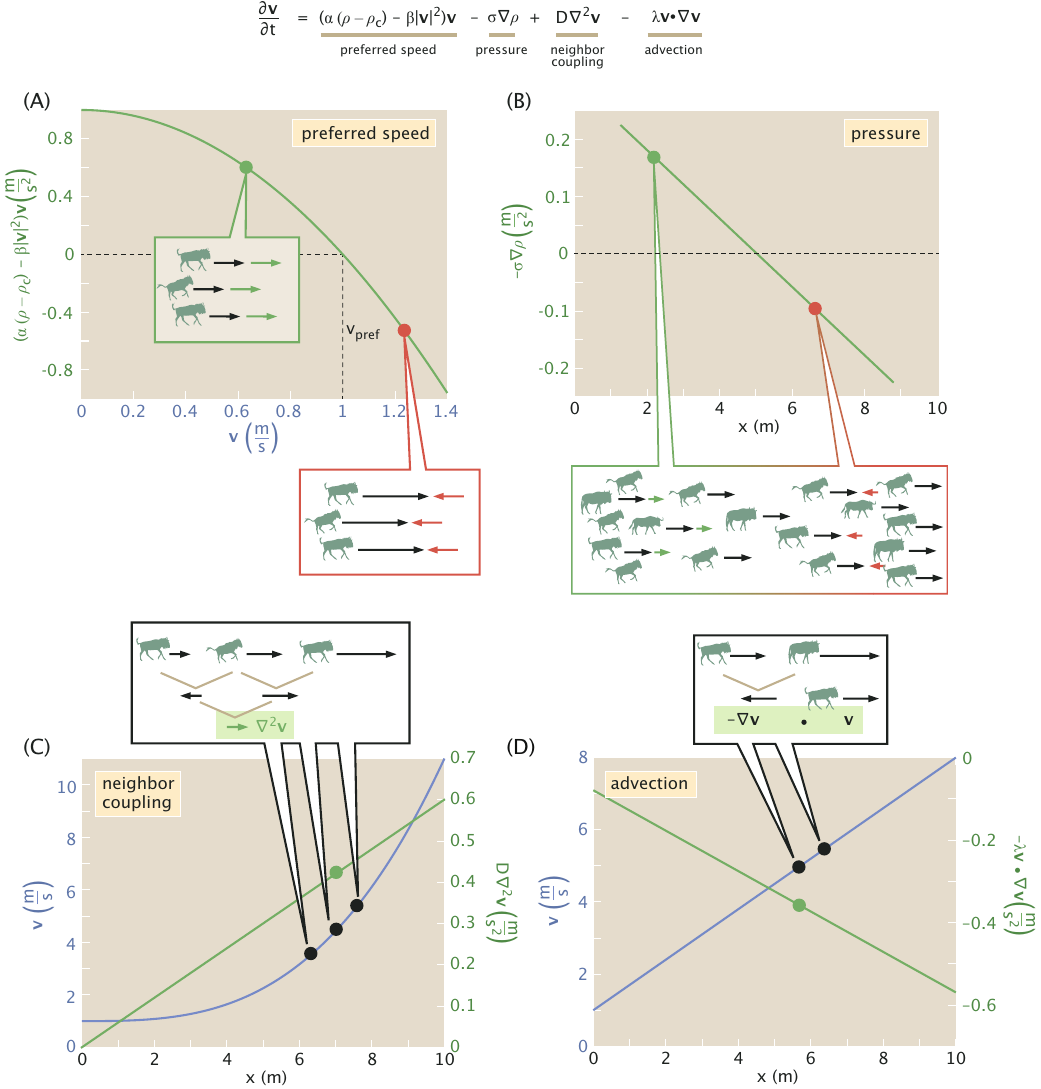}}
\caption{Building intuition for the physics of the velocity field in the Toner-Tu theory.  The different terms in the Toner-Tu
analysis are represented graphically for a one-dimensional herd.  We can think of each term as providing an update to the current velocity. In each case, an example of a simple velocity profile (blue) or density profile (graphic in B) is shown alongside the corresponding velocity update (green). (A) The preferred speed term increases the velocity of wildebeest that are moving too slow and decreases the velocity of wildebeest that are moving too fast. (B) The pressure term punishes gradients in density, adjusting velocity in order to flatten these gradients. Given mass conservation in a finite system, the pressure term alone would lead to a steady state of uniform density $\rho = \rho_{0}$. (C) The neighbor coupling term captures the velocity adjustment made by a wildebeest to better match its neighbors, smoothing out differences in velocity. A given wildebeest (middle black dot) adjusts its velocity by an amount represented by the green arrow: an average of the difference between its velocity and its two neighbors' velocities. This averaging of two differences is mathematically analogous to taking a local second derivative or local Laplacian, $\nabla^2 \mathrm{{\bf v}}$. (D) Finally, the advection term ensures that the filament velocity field is swept along according to its own velocities, just as fluid velocity is self-advected in the Navier-Stokes equations. Mathematically, the velocity adjustment needed to ensure velocity advection is a function of the spatial gradient of velocity ($-\nabla {\bf v}$, how mismatched in velocity nearby wildebeest are) and the velocity itself (${\bf v}$, how fast that mismatch is carried along by the herd).
\label{fig:FlockingIntuition}}
\end{figure}

The first step in the development of continuum theories of active matter involves selecting
the relevant field variables.   Following the classic work of Toner and Tu~\cite{Toner1995,Toner1998} and inspired by observations of
herds of land animals like wildebeest, we consider a two-dimensional
density field $\rho({\bf r},t)$ and a corresponding two-dimensional  velocity field ${\bf v}({\bf r},t)$, which captures both the orientation and speed of a polar agent. 
Throughout this study, wildebeest herds will serve as inspiration and example,
although in a sense the term ``wildebeest" is our shorthand for a ``self-propelled polar agent"
 at any length scale.
 We also note that the Toner-Tu model used here describes \textit{dry} systems, in which momentum is not conserved and hydrodynamic coupling between the agents is negligible relative to frictional drag. 
With variables $\rho$ and ${\bf v}$ in hand, our next step is to write the partial differential equations
that describe the spatiotemporal evolution of those field variables.
The first governing equation is the continuity equation given by 
\begin{equation}
{\partial \rho \over \partial t}+{\partial (\rho v_i)\over \partial x_i}=0,
\end{equation}
which allows modulations in the density field but enforces mass conservation, forbidding wildebeest birth or death in the midst of herding phenomena. Note that we are using the Einstein summation convention, which tells us to sum over
all repeated indices. For example,  ${\bf a} \cdot {\bf b} = \sum_{i=1}^3 a_ib_i = a_ib_i$. 
The second governing equation is a minimal representation of
the dynamics of the ${\bf v}$ field offered by Toner and Tu \cite{Toner1995} and is given by
\begin{equation}
{\partial v_i \over \partial t}=
[\alpha (\rho - \rho_c) - \beta v_jv_j]v_i -\sigma {\partial \rho \over \partial x_i}+D \nabla^2 v_i 
- \lambda v_j {\partial v_i \over \partial x_j},
\label{eqn:TonerTuClassic1}
\end{equation}
where $\rho_{c}$ is the critical density above which the herd moves coherently, the ratio of $\alpha(\rho-\rho_{c})$ and $\beta$ sets the wildebeest mean-field speed $v_{\mbox{pref}} = \sqrt{\alpha \left(\rho-\rho_{c}\right)/\beta}$, the term $\sigma ~\partial \rho / \partial x_i$ provides an effective pressure, $D$ tunes wildebeest alignment and speed matching with neighbors, and $\lambda$ tunes velocity self-advection. 
 We can conceptualize each term as an ``update rule" that computes incremental changes in the velocity vector at each point in space and at every step in time.
 In Figure~\ref{fig:FlockingIntuition} and in the remainder of this subsection, we seek to provide an intuitive interpretation of each term and its contribution to updating the velocity field.  
 
The first term on the right hand side of eqn.~\ref{eqn:TonerTuClassic1} , the preferred speed term, pushes the velocity magnitude toward the characteristic speed of a wildebeest, $v_{\mbox{\footnotesize{pref}}}$ \cite{Toner2018}. 
 The second is based on an equation of state originally used
 by Toner and Tu to relate density and pressure \cite{Toner1998}. This pressure term punishes gradients in density, adjusting velocity to flatten the density field. 
 The third, the neighbor coupling term, provides a smoothing or diffusion of velocity (orientation and speed) that reflects coordination between nearby wildebeest.  Interestingly, if wildebeest adopt the rule that a given wildebeest averages the difference between its own velocity and that of its neighbors, mathematically, the result is a Laplacian term like that seen in the minimal model~\cite{Toner2018}.
 The final term has analogy to the gradient component of the material time derivative in the Navier-Stokes equations. In essence, the velocity field advects itself; wildebeest move along in a direction dictated by their orientation and velocity, and they bring that orientation and velocity with them. In the case of pure velocity self-advection, $\lambda$ = 1, but this is not necessarily the case for active flocks \cite{Toner2018, Dadhichi2020}. 
In the case of wildebeest herds, we can conceptualize this effect by considering that $\lambda$ = 1 + $\xi$, where $\xi$ reflects a behavioral response to gradients in velocity; for example, wildebeest may resist running quickly into a steep gradient of decreasing velocity, and may slow down.

For a full pedagogical derivation of the Toner-Tu model, we recommend a series of lectures by John Toner \cite{Toner2018} in which he uses symmetry arguments to infer what terms should be kept in a complete continuum description of herding and also provides intuitive arguments about what these terms mean.

\subsection{Formulating the Theory for Arbitrary Curved Surfaces}

\begin{figure}
\centering{\includegraphics[width=5.0truein]{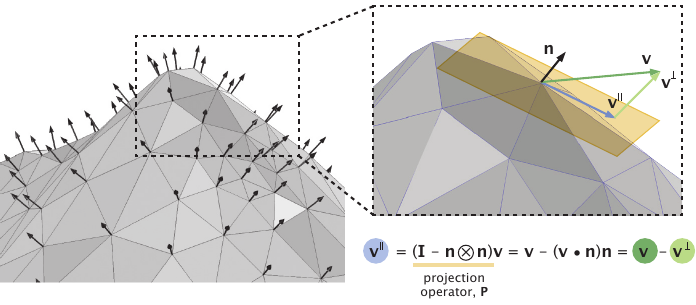}}
\caption{Illustration of the projection operator essential to the general surface formulation and finite element treatment of the Toner-Tu 
equations on curved surfaces.  The surface of
interest is represented by a collection of nodes (vertices of triangles) and a field of surface normal vectors, ${\bf n}$ (black arrows). An arbitrary vector ${\bf v}$ is projected onto the tangent plane (shown in gold),
permitting a decomposition of the form ${\bf v}= {\bf v}^{\parallel}+{\bf v}^{\perp}$.
\label{fig:ProjectionApproach}}
\end{figure}

Inspired by a desire to make contact with phenomena of the natural world, like wildebeest navigating an undulating landscape, sheep flocks crossing hilly pastures, and the surfaces flows of flocking actin we study in reference \cite{Hueschen2022c}, we sought to solve the minimal Toner-Tu theory presented above on complex and asymmetric surface geometries.
We consider here flocking on non-deformable surfaces, but we refer the interested reader to earlier work on surface hydrodynamics in the completely general case
in which the surface itself can evolve over time~\cite{Voigt2017, Sahu2017, Jankuhn2018,TorresSanchez2019}. 
Predicting flocking or herding behavior on arbitrary surfaces requires us
to reformulate the theory in a more general way that accounts
for curvature.  Instead of basing our formulation on a parameterized surface
and the intrinsic differential geometry tools that this approach
licenses, as done in beautiful earlier work~\cite{Shankar2017}, from the outset we have in mind arbitrary surfaces that can be represented using finite element meshes and
described by a field of local normal vectors, ${\bf n}$.
Our choice and use of this extrinsic differential geometry and finite element method approach was aided by the work of many, including refs. \cite{Sahu2017, Jankuhn2018, Mietke2019, Mietke2018, Nestler2018,  Fries2018, Nestler2019, Nitschke2019, Rangamani2013} and chapter 3 of the supplemental material for \cite{Takatori2020}.
The finite element setting permits us the convenience of
carrying out the mathematics  in the
full three-dimensional setting of $\mathbb{R}^3$, while using our knowledge of the normal vectors
everywhere on the surface of interest to project our governing equations onto
the surface. While velocity is described by the 3-dimensional vector {\bf v}, both {\bf v} and the scalar $\rho$ are defined only on the surface. At every point on the surface, derivatives evaluated in the usual
$\mathbb{R}^3$ way are projected onto the tangent plane using the local normal.  For an insightful description of this extrinsic differential geometry approach to handling curved surfaces and its mathematical equivalence to the intrinsic differential geometry strategy, we recommend chapters 22 and 23 of Needham~\cite{Needham2020}.  Central to the extrinsic geometry approach is the projection operator, defined as
\begin{equation}
\mbox{projection onto tangent plane}={\bf P} =  {\bf I}- {\bf n} \otimes {\bf n},
\label{eqn:ProjectionOperator}
\end{equation}
where ${\bf I}$ is the identity matrix and ${\bf n} \otimes {\bf n}$ is the outer product of
the surface normal vector as shown in Figure~\ref{fig:ProjectionApproach}.
To make sense of this expression mathematically,
we recall that  the outer product  ${\bf n} \otimes {\bf n}$ is defined through its action on a vector ${\bf v}$ as
\begin{equation}
({\bf n} \otimes {\bf n}) {\bf v} = {\bf n} ({\bf n} \cdot {\bf v}).
\end{equation}
We can write ${\bf v}= {\bf v}^{\parallel}+{\bf v}^{\perp}$,
where ${\bf v}^{\parallel}$ is the component of ${\bf v}$ in the tangent plane
of the surface and ${\bf v}^{\perp}$ is normal to the surface. 
The action of  ${\bf I}- {\bf n} \otimes {\bf n}$ on a vector
${\bf v}$ is given by
\begin{equation}
 ({\bf I}- {\bf n} \otimes {\bf n}) {\bf v}= {\bf v} - {\bf n} ({\bf n} \cdot {\bf v})= {\bf v}- {\bf v}^{\perp}= {\bf v}^{\parallel},
 \end{equation}
 as illustrated in Figure~\ref{fig:ProjectionApproach}.
We can write the projection operator in component form as
\begin{equation}
P_{ij}=\delta_{ij}-n_in_j,
\label{eqn:ProjectionOperatorIndicial}
\end{equation}
recalling that the normal vector to the surface is given by ${\bf n}=(n_1,n_2,n_3)$,
or in full  component form  as
\begin{equation}
{\bf P}=\left[\begin{array}{ccc}
1-n_{1}^{2} & -n_{1} n_{2} & -n_{1} n_{3} \\
-n_{1} n_{2} & 1-n_{2}^{2} & -n_{2} n_{3} \\
-n_{1} n_{3} & -n_{2} n_{3} & 1-n_{3}^{2}
\end{array}\right].
\end{equation}
Using the projection operator allows us to perform calculations in
the ordinary three-dimensional space within which the surface of interest is embedded, but
then to pick off only the pieces of the resulting vectors that live within the surface.

In the remainder of this section, we examine each term in the minimal Toner-Tu equations and translate it into its projected form. In Appendix Section \ref{Section:VectorTensor}, we define the projected surface form of calculus operators as an additional reference for the reader.
First, to modify the preferred speed term to its curved-surface implementation,
we note that it is now ${\bf v}^{\parallel}$, the in-plane velocity, that has a privileged
magnitude. This magnitude is imposed through the condition
\begin{equation}
\mbox{preferred speed term}=[\alpha (\rho - \rho_c) - \beta v_j^{\parallel} v_j^{\parallel}]v_i^{\parallel}.
\label{eqn:VelocitySelectionTerm}
\end{equation} 
If $|{\bf v}^{\parallel}|$ is either larger or smaller than
the privileged value $v_{\mbox{\footnotesize{pref}}}=\sqrt{\alpha(\rho-\rho_c)/\beta}$, this term will
adjust the velocity towards that steady-state magnitude.

 Next, we consider the pressure term, whose physical origin comes from a model of pressure  in powers of density of the form introduced by Toner and Tu~\cite{Toner1995},
\begin{equation}
P(\rho)=\sum_{n=1}^{\infty} \sigma_n (\rho-\rho_0)^n
\end{equation}
where $\rho_{0}$ is the mean density.
 In the minimal Toner-Tu theory we adopt here, only the first order term in that expansion  is kept,  with the notational simplification that $\sigma_1=\sigma$, resulting in
\begin{equation}
\mbox{pressure term}= -{\partial P \over \partial x_i} = -\sigma {\partial \rho \over \partial x_i}.
\end{equation}
We note that while $\rho_{0}$ is not explicitly present in the gradient of the first order pressure term, for the case considered here of a finite surface on which total density is conserved, this pressure term effectively maintains a density range centered around the mean density $\rho_{0}$ established by our choice of initial condition.
The curved-space version of the pressure term requires projecting the full 3D gradient of the density onto the tangent plane, using the projection operator defined in eqn.~\ref{eqn:ProjectionOperator}.  We follow Jankuhn {\it et al.}~\cite{Jankuhn2018} in introducing
the notation $\nabla_{\Gamma} $ for the projected surface gradient operator, where the subscript $\Gamma$ indicates that the gradient is evaluated on the surface of interest.
Mathematically, this amounts to computing  
\begin{equation}
\nabla_{\Gamma} \rho = ({\bf I}-{\bf n} \otimes {\bf n}) \nabla \rho,
\end{equation}
where $ \nabla$  is the ordinary, three-dimensional gradient in Cartesian coordinates.
This can be rewritten in component form as
\begin{equation}
\begin{aligned}
(\nabla_{\Gamma} \rho)_i = \left[({\bf I}-{\bf n} \otimes {\bf n}) \nabla \rho\right]_i =\left(\delta_{i j}-n_{i} n_{j}\right) \frac{\partial \rho}{\partial x_{j}}
=\frac{\partial \rho}{\partial x_{i}}-n_{i} n_{j} \frac{\partial \rho}{\partial x_{j}}.
\end{aligned}
\label{eqn:PressureTerm}
\end{equation}

The next term in the Toner-Tu equations that we consider in its curved-space  format
is the advection term in eqn.~\ref{eqn:TonerTuClassic1}, namely,
\begin{equation}
\mbox{advection term}=  -\lambda v_j {\partial v_i \over \partial x_j}.
\end{equation}
In direct notation, the curved-space version of this term has the form
\begin{equation}
\mbox{advection term}=  -\lambda ( {\bf v}^{\parallel} \cdot \nabla_{\Gamma})  {\bf v}^{\parallel},
\end{equation}
which involves the curved-space version of the velocity gradient tensor. In Appendix Section \ref{Section:VectorTensor}, we describe how to compute this tensor, which stated simply is 
\begin{equation}
\nabla_{\Gamma} {\bf v}^{\parallel}= {\bf P} ( \nabla {\bf v}^{\parallel} ) {\bf P},
\end{equation}
where {\bf P} is the projection operator defined in eqn. \ref{eqn:ProjectionOperator}. In indicial notation, this leads to the result
\begin{equation}
(\nabla_{\Gamma} {\bf v}^{\parallel})_{ij}=  [  {\bf P} ( \nabla {\bf v}^{\parallel} ) {\bf P} ]_{ij}  =
P_{ik} {\partial v_k^{\parallel} \over \partial x_l} P_{lj}.
\end{equation}
Invoking the definition of the projection operator from eqn. \ref{eqn:ProjectionOperatorIndicial}, this expression simplifies to the result
\begin{equation}
(\nabla_{\Gamma} {\bf v}^{\parallel})_{ij}= [  {\bf P} ( \nabla {\bf v}^{\parallel} ) {\bf P} ]_{ij} =
\left(\frac{\partial v_{i}^{\parallel}}{\partial x_{j}}-n_ln_{j} \frac{\partial v_{i}^{\parallel}}{\partial x_{l}}\right) - n_in_k \left(\frac{\partial v_{k}^{\parallel}}{\partial x_{j}}-n_ln_{j} \frac{\partial v_{k}^{\parallel}}{\partial x_{l}} \right).
\label{eqn:VelocityGradientCurved}
\end{equation}
Thus, the curved-space advection term can be written as
\begin{equation}
-\lambda   [  {\bf P} ( \nabla {\bf v}^{\parallel} ) {\bf P} ]_{ij}  v_j^{\parallel}=-\lambda \left[
\left(\frac{\partial v_{i}^{\parallel}}{\partial x_{j}}-n_ln_{j} \frac{\partial v_{i}^{\parallel}}{\partial x_{l}}\right) - n_in_k \left(\frac{\partial v_{k}^{\parallel}}{\partial x_{j}}-n_ln_{j} \frac{\partial v_{k}^{\parallel}}{\partial x_{l}} \right)\right] v_j^{\parallel}.
\label{eqn:AdvectionTerm}
\end{equation}

With these examples of tangent-plane calculus established, we now turn to
the most tricky of the terms in the Toner-Tu equations, 
the surface-projected version of the neighbor coupling term, $D \nabla^2 {\bf v}$.
 First, we recall that
the Laplacian of a vector field in normal 3D Cartesian
space is defined as $\nabla^2 {\bf v} = \nabla \cdot (\nabla {\bf v})$.
Further, we note that $\nabla {\bf v}$ is itself a tensor.
As shown by Jankuhn {\it et al.}~\cite{Jankuhn2018}, the surface-projected version of the Laplacian term (see their  eqn.~3.16 for the surface Navier-Stokes equations for the tangential velocity on
a stationary surface), is therefore given by
  \begin{equation}
\underbrace{\nabla^2{\bf v}}_{\mbox{flat Toner-Tu}}  = \underbrace{\mbox{div}( \nabla {\bf v})}_{\mbox{in the plane}} ~\longrightarrow ~~~\underbrace{{\bf P} \mbox{div}_{\Gamma}({\bf G}({\bf v}^{\parallel}))}_{\mbox{curved surface}}.
 \end{equation}
To begin to unpack this expression, we note that ${\bf P}$ is the projection operator defined in eqn.~\ref{eqn:ProjectionOperator} and that the tensor ${\bf G}$ is the surface velocity gradient,
\begin{equation}
 {\bf G}({\bf v}^{\parallel})= \nabla_{\Gamma} {\bf v}^{\parallel},
 \label{eqn:GDefined}
 \end{equation}
already presented in eqn.~\ref{eqn:VelocityGradientCurved}
and repeated here in component form as
 \begin{equation}
 G_{ij}=\left(\frac{\partial v_{i}^{\parallel}}{\partial x_{j}}-n_ln_{j} \frac{\partial v_{i}^{\parallel}}{\partial x_{l}}\right) - n_in_k \left(\frac{\partial v_{k}^{\parallel}}{\partial x_{j}}-n_ln_{j} \frac{\partial v_{k}^{\parallel}}{\partial x_{l}} \right).
 \label{eqn:GMatrix}
 \end{equation}
We note that Jankuhn {\it et al.}~\cite{Jankuhn2018}
use a symmetrized version of the velocity gradient since they are deriving the surface
versions of the Navier-Stokes equations and are thus taking the divergence of a stress.
For the Toner-Tu case of interest here, the ``psychological viscosity'' that comes from 
neighboring wildebeest in the herd comparing their velocities is equivalent to the divergence of
the velocity gradient itself.
We then invoke the definition of the surface-projected divergence of a tensor ${\bf G}$ presented in Appendix Section \ref{Section:VectorTensor}, giving
rise to a vector of the form 
 \begin{equation}
\operatorname{div}_{\Gamma}\left({\bf G} \right)_l =\frac{\partial G_{l j}}{\partial x_{j}}-n_{j}n_k \frac{\partial G_{l j}}{\partial x_{k}}.
\label{eqn:DivergenceTensorIndicial2}
\end{equation}
We also introduce the shorthand notation for this divergence as
\begin{equation}
\left( {\partial G_{lj} \over \partial x_j}\right)_{\Gamma}=\frac{\partial G_{l j}}{\partial x_{j}}-n_{j}n_k \frac{\partial G_{l j}}{\partial x_{k}}
\label{eqn:DivergenceTensorIndicial2Shorthand}
\end{equation}
to simplify some of the complex expressions to follow.
Finally,  we can write the $i^{th}$ component of the surface version of the Toner-Tu term $D \nabla^2 {\bf v}$ in indicial notation as
 \begin{equation}
D [{\bf P} \mbox{div}_{\Gamma}({\bf G}({\bf v}^{\parallel}))]_i = D P_{il} \Big(\frac{\partial G_{l j}}{\partial x_{j}}-n_{j}n_k \frac{\partial G_{l j}}{\partial x_{k}} \Big).
 \label{eqn:LaplacianTermProj}
 \end{equation}

We now assemble
the results of eqns.~\ref{eqn:VelocitySelectionTerm}, ~\ref{eqn:PressureTerm},
~\ref{eqn:AdvectionTerm} and ~\ref{eqn:LaplacianTermProj} to construct a complete curved-surface formulation of the minimal
Toner-Tu equations.  Specifically, we have
\begin{equation}
{\partial  v_i^{\parallel} \over \partial t}=[\alpha (\rho - \rho_c) - \beta v_j^{\parallel} v_j^{\parallel}]v_i^{\parallel}
-\sigma \left(\frac{\partial \rho}{\partial x_{i}}- n_{i} n_{j} \frac{\partial \rho}{\partial x_{j}}\right) 
+D P_{il} \left(\frac{\partial G_{l j}}{\partial x_{j}}-n_{j}n_k \frac{\partial G_{l j}}{\partial x_{k}} \right)
-\lambda  [  {\bf P} ( \nabla {\bf v}^{\parallel} ) {\bf P} ]_{ij}  v_j^{\parallel},
  \label{eqn:TonerTuCurvedSpace}
\end{equation}
which can be streamlined to the alternative form
\begin{equation}
{\partial  v_i^{\parallel} \over \partial t}=[\alpha (\rho - \rho_c) - \beta v_j^{\parallel} v_j^{\parallel}]v_i^{\parallel}
-\sigma \left(\frac{\partial \rho}{\partial x_{i}}\right)_{\Gamma} +D P_{il} \left(\frac{\partial G_{l j}}{\partial x_{j}}\right)_{\Gamma} -\lambda v_j^{\parallel} \left[\left({\partial  v_i^{\parallel}\over \partial x_j}\right)_{\Gamma}-n_in_k \left({\partial  v_k^{\parallel}\over \partial x_j}\right)_{\Gamma} \right].
 \label{eqn:TonerTuCurvedSpace2}
\end{equation}
Similarly, the curved-space formulation of the governing equation for density can be written as
\begin{equation}
\frac{\partial \rho}{\partial t}=-\frac{\partial (\rho v_i^{\parallel})}{\partial x_i} + n_i n_j \frac{\partial (\rho v_i^{\parallel})}{\partial x_j} = - \left({\partial \rho v_i^{\parallel} \over \partial x_i}\right)_{\Gamma}.
\label{eqn:ContinuityProjected}
\end{equation}
We now have a complete formulation of our minimal Toner-Tu equations for the general surface context, requiring only a description of the surface in the language of normal vectors.
We next turn to the implementation of this general surface formulation in a fashion consistent with
finite element treatments on arbitrary surfaces.  

\section{Formulating the Surface Toner-Tu Equations for Numerical Implementation}
\label{Section:FiniteElement}

Numerically solving active matter equations on complex
surfaces relevant to the living world presents a practical challenge. It precludes the differential geometric formalism used to describe parameterized surfaces, which would lead to equations featuring covariant derivatives (e.g., eqn.~\ref{eqn:FullTonerTuCurved} in Section \ref{Section:TonerTuForParameterizedSurfaces}).
In the finite element setting,
the surface of interest is represented by a collection of nodes and corresponding surface normals, as shown in Figure \ref{fig:ProjectionApproach}.
Using these surface normals, we perform surface projections of the full 3-space derivatives following Jankuhn {\it et al.}~\cite{Jankuhn2018}.
In the previous section, we presented a formal statement in eqn.~\ref{eqn:TonerTuCurvedSpace} for handling the minimal Toner-Tu model on an arbitrary surface, using these surface normals.  We now recast those curved-surface
equations once more, in a fashion consonant with a  finite element method solver. 
To formulate the equations in an expression convenient for
the finite element method,  we aim to rewrite the Toner-Tu equations in the form
 \begin{equation}
 \frac{\partial \mathbf{v}}{\partial t}+\nabla 
\cdot {\bf J}^{(v)}={\bf f}^{(v)} .
\label{eqn:FEMFormulationTonerTu}
\end{equation}
 Here,  the whole formulation comes
down to the definitions of ${\bf J}^{(v)}$ and ${\bf f}^{(v)} $. 
The flux-like quantity ${\bf J}^{(v)}$ is a $3 \times 3$ matrix defined such
that 
\begin{equation}
\nabla 
\cdot {\bf J}^{(v)}=( {\partial \over \partial x_1}, {\partial \over \partial x_2}, {\partial \over \partial x_3})
\left[\begin{array}{lll}
J_{11} & J_{12} & J_{13} \\
J_{21} & J_{22} & J_{23} \\
J_{31} & J_{32} & J_{33}
\end{array}\right],
\end{equation}
recalling that the divergence of a 2nd rank tensor is a vector.
In indicial notation, this can be written as
\begin{equation}
(\nabla 
\cdot {\bf J}^{(v)})_i = {\partial J_{ji}^{(v)} \over \partial x_j}.
\end{equation}
We need to define the components of the tensor ${\bf J}^{(v)}$ such that
they yield the correct Toner-Tu terms.  As we will see below,
${\bf J}^{(v)}$ is not symmetric despite its superficial resemblance to a stress tensor.
To set notation and to make sure that the strategy is clear, we begin by demonstrating how
to implement the flat-space version of the Toner-Tu equations 
in a form consistent with eqn.~\ref{eqn:TonerTuClassic1}.   If
we define the force term ${\bf f}^{(v)} $ as 
\begin{equation}
f_i^{(v)}  =
 [\alpha (\rho - \rho_c) - \beta v_jv_j]v_i - \sigma {\partial \rho \over \partial x_i} -\lambda v_j {\partial v_i \over \partial x_j},
\label{eqn:ForcingCOMSOL}
\end{equation}
then the remaining terms are captured if 
we define
the tensor ${\bf J}^{(v)}$  as
\begin{equation}
{\bf J}^{(v)}=\left[\begin{array}{ccc}
-D  {\partial v_1 \over \partial x_1}&  -D  {\partial v_2 \over \partial x_1} & -D {\partial v_3 \over \partial  x_1}\\ [2ex]
-D {\partial v_1 \over \partial x_2}& -D {\partial v_2 \over \partial x_2} & -D {\partial v_3 \over \partial  x_2}\\[2ex]
-D  {\partial v_1 \over \partial x_3}& -D {\partial v_2 \over \partial x_3}&  -D {\partial v_3 \over \partial  x_3}
\end{array}\right].
\label{eqn:FEMflux}
\end{equation}
Considering the 1-component of velocity, we see that
\begin{equation}
(\nabla 
\cdot {\bf J}^{(v)} )_1= {\partial J_{j1} \over \partial x_j}= 
 {\partial J_{11} \over \partial x_1}+ {\partial J_{21} \over \partial x_2}
 + {\partial J_{31} \over \partial x_3}.
 \end{equation}
 Plugging in the components of ${\bf J}^{(v)}$ gives the result
 \begin{equation}
(\nabla 
\cdot {\bf J}^{(v)} )_1= {\partial \over \partial x_1} \left( -D  {\partial v_1 \over \partial x_1}\right)
+{\partial \over \partial x_2}  \left( -D {\partial v_1 \over \partial x_2}\right)
+{\partial \over \partial x_3}  \left( -D {\partial v_1 \over \partial x_3}\right). 
\end{equation}
Evaluating these derivatives leads to the result
 \begin{equation}
(\nabla 
\cdot {\bf J}^{(v)} )_1= - D{\partial^2 v_1 \over \partial x_1^2}-D{\partial^2 v_1 \over \partial x_2^2}-D{\partial^2 v_1 \over \partial x_3^2},
 \label{eqn:TonerTuDivergenceFlat}
 \end{equation}
 as expected from the original Toner-Tu equations.  
Combining the contributions from eqns.~\ref{eqn:ForcingCOMSOL} and ~\ref{eqn:FEMflux} in the form $ \partial \mathbf{v} / \partial t +\nabla 
\cdot {\bf J}^{(v)}={\bf f}^{(v)}$, 
we recover eqn.~\ref{eqn:TonerTuClassic1} precisely
 as we set out to do.   By defining the quantities ${\bf J}^{(v)}$ and ${\bf f}^{(v)} $, we have successfully reframed the flat-space Toner-Tu equations in a format that will be conveniently implemented
 in the finite element setting.
We now need
to tackle the more demanding formulation for 
an arbitrary curved surface. 

Abstractly, our finite element version of the curved-space Toner-Tu equations is
written as
\begin{equation}
{\partial {\bf v} \over \partial t} =- \operatorname{div}_{\Gamma} {\bf J}^{(v)}+ {\bf f}^{(v)} ,
\end{equation}
where $\operatorname{div}_{\Gamma}$ is the surface projected version of the
divergence introduced in eqn.~\ref{eqn:DivergenceTensorIndicial2}. 
We now introduce the definition
\begin{equation}
J_{j i}^{(v)}= -D P_{il} G_{lj},
\label{eqn:FEMfluxCurvedSpace}
\end{equation}
where we recall ${\bf P}$ from eqn.~\ref{eqn:ProjectionOperator} and ${\bf G}$ from
eqn.~\ref{eqn:GDefined}.
Given this definition, we can attempt to write our Toner-Tu equations 
as
\begin{equation}
{\partial v_i \over \partial t} =-D \left({\partial (-P_{il} G_{lj}) \over \partial x_j}\right)_{\Gamma}  + f_i^{(v)},
\end{equation}
where we use the shorthand notation introduced in eqn.~\ref{eqn:DivergenceTensorIndicial2Shorthand}.
This can be rewritten as
\begin{equation}
{\partial v_i \over \partial t} = D \underbrace{P_{i l} \left(\frac{\partial G_{lj}}{\partial x_{j}}\right)_{\Gamma}}_{{\bf P} \mbox{div}_{\Gamma}(\nabla_{\Gamma} {\bf v}^{\parallel})}+~D\underbrace{\left(\frac{\partial P_{i l}}{\partial x_{j}}\right)_{\Gamma}G_{lj}}_{\mbox{unwanted term}}+ f_i^{(v)}.
\end{equation}
Unfortunately, as written, these equations contain an extra term that is not present in the Toner-Tu formulation.  
To remove this unwanted extra term, we must introduce a new term in the force ${\bf f}^{(v)}$ that subtracts off the unwanted term,
\begin{equation}
f_i^{\mbox{fict}}= -D\left(\frac{\partial P_{i l}}{\partial x_{j}}\right)_{\Gamma} G_{lj}.
\end{equation}
We view this as a mathematical trick that allows us to use a formalism convenient for finite element method analysis and have
not sought to find a ``physical interpretation'' of this force in the way that ideas
such as the Coriolis force arises in mechanics.
In this case, our {\it real}  Toner-Tu equations can be written as 
\begin{equation}
{\partial v_i \over \partial t} =D \underbrace{P_{i l} \left(\frac{\partial G_{lj}}{\partial x_{j}}\right)_{\Gamma}}_{{\bf P} \mbox{div}_{\Gamma}(\nabla_{\Gamma} {\bf v}^{\parallel})}+ ~D\underbrace{\left(\frac{\partial P_{i l}}{\partial x_{j}}\right)_{\Gamma}G_{lj}}_{\mbox{unwanted term}}+ f_i^{(v)}+ f_i^{\mbox{\mbox{fict}}}.
\label{eqn:TonerTuFEMComplete}
\end{equation}
We now have precisely the equations we want, cast in the form we will use in
the finite element setting.

As we saw above, the force term ${\bf f}^{(v)}+{\bf f}^{\mbox{\mbox{fict}}}$ needs to account for those terms that are not present in $\nabla 
\cdot {\bf J}^{(v)}$ and to subtract off terms that are present in
$\nabla 
\cdot {\bf J}^{(v)}$ but unwanted.  To that end, the 1-component of ${\bf f}^{(v)}+{\bf f}^{\mbox{\mbox{fict}}}$   takes the form 
\begin{equation}
\begin{aligned}
f_1^{(v)}+ f_1^{\mbox{\mbox{fict}}} & =
 \Big(\alpha (\rho-\rho_c)- \beta ((v_1^{\parallel})^2 + (v_2^{\parallel})^2+(v_3^{\parallel})^2) \Big)v_1^{\parallel} \\
&-\sigma \left({\partial \rho \over \partial x_1}\right)_{\Gamma} -\lambda \Big(v_1^{\parallel} ~{\partial  v_1^{\parallel} \over \partial x_1}+v_2^{\parallel} ~{\partial  v_1^{\parallel} \over \partial x_2}+
v_3^{\parallel} ~{\partial  v_1^{\parallel} \over \partial x_3}\\
&-v_1^{\parallel}n_1n_1 ~{\partial  v_1^{\parallel} \over \partial x_1}
-v_1^{\parallel}n_1n_2 ~{\partial  v_2^{\parallel} \over \partial x_1}
-v_1^{\parallel}n_1n_3 ~{\partial  v_3^{\parallel} \over \partial x_1}\\
&-v_2^{\parallel}n_1n_1 ~{\partial  v_1^{\parallel} \over \partial x_2}
-v_2^{\parallel}n_1n_2 ~{\partial  v_2^{\parallel} \over \partial x_2}
-v_2^{\parallel}n_1n_3 ~{\partial  v_3^{\parallel} \over \partial x_2}\\
&-v_3^{\parallel}n_1n_1 ~{\partial  v_1^{\parallel} \over \partial x_3}
-v_3^{\parallel}n_1n_2 ~{\partial  v_2^{\parallel} \over \partial x_3}
-v_3^{\parallel}n_1n_3 ~{\partial  v_3^{\parallel} \over \partial x_3}\Big)\\
&-D \left( \left({\partial P_{11} \over \partial x_1}\right)_{\Gamma} G_{11}+ \left({\partial P_{11} \over \partial x_2}\right)_{\Gamma} G_{12}+\left({\partial P_{11} \over \partial x_3}\right)_{\Gamma} G_{13}\right) \\
&-D\left(\left({\partial P_{12} \over \partial x_1}\right)_{\Gamma} G_{21}+ \left({\partial P_{12} \over \partial x_2}\right)_{\Gamma}G_{22}+\left({\partial P_{12} \over \partial x_3}\right)_{\Gamma}G_{23}\right)\\
&-D \left(\left({\partial P_{13} \over \partial x_1}\right)_{\Gamma}G_{31}+ \left({\partial P_{13} \over \partial x_2}\right)_{\Gamma}G_{32}+\left({\partial P_{13} \over \partial x_3}\right)_{\Gamma}G_{33}\right).
\end{aligned}
\label{eqn:FEMforcing1}
\end{equation}
Note that 
the $\alpha$ and $\beta$ terms capture the preferred speed
contribution,  the $\sigma$ term captures the pressure, 
the terms multiplied by $\lambda$ capture the advection contribution 
to the Toner-Tu equations,
and the final terms involving ${\bf P}$ and ${\bf G}$ subtract
off the fictitious force.  We can repeat a similar analysis for the 2- and 3- components of the force as
\begin{equation}
\begin{aligned}
f_2^{(v)}+ f_2^{\mbox{\mbox{fict}}}  & =
 \Big(\alpha (\rho-\rho_c)- \beta ((v_1^{\parallel})^2 + (v_2^{\parallel})^2+(v_3^{\parallel})^2) \Big)v_2^{\parallel} \\
&- \sigma \left({\partial \rho \over \partial x_2}\right)_{\Gamma} -\lambda \Big(v_1^{\parallel} ~{\partial  v_2^{\parallel} \over \partial x_1}+v_2^{\parallel} ~{\partial  v_2^{\parallel} \over \partial x_2}+
v_3^{\parallel} ~{\partial  v_2^{\parallel} \over \partial x_3} \\
&-v_1^{\parallel}n_2n_1 ~{\partial  v_1^{\parallel} \over \partial x_1}
-v_1^{\parallel}n_2n_2 ~{\partial  v_2^{\parallel} \over \partial x_1}
-v_1^{\parallel}n_2n_3 ~{\partial  v_3^{\parallel} \over \partial x_1}\\
&-v_2^{\parallel}n_2n_1 ~{\partial  v_1^{\parallel} \over \partial x_2}
-v_2^{\parallel}n_2n_2 ~{\partial  v_2^{\parallel} \over \partial x_2}
-v_2^{\parallel}n_2n_3 ~{\partial  v_3^{\parallel} \over \partial x_2}\\
&-v_3^{\parallel}n_2n_1 ~{\partial  v_1^{\parallel} \over \partial x_3}
-v_3^{\parallel}n_2n_2 ~{\partial  v_2^{\parallel} \over \partial x_3}
-v_3^{\parallel}n_2n_3 ~{\partial  v_3^{\parallel} \over \partial x_3}\Big)\\
&-D \left( \left({\partial P_{21} \over \partial x_1}\right)_{\Gamma} G_{11}+ \left({\partial P_{21} \over \partial x_2}\right)_{\Gamma} G_{12}+\left({\partial P_{21} \over \partial x_3}\right)_{\Gamma} G_{13}\right) \\
&-D\left(\left({\partial P_{22} \over \partial x_1}\right)_{\Gamma} G_{21}+ \left({\partial P_{22} \over \partial x_2}\right)_{\Gamma}G_{22}+\left({\partial P_{22} \over \partial x_3}\right)_{\Gamma}G_{23}\right)\\
&-D \left(\left({\partial P_{23} \over \partial x_1}\right)_{\Gamma}G_{31}+ \left({\partial P_{23} \over \partial x_2}\right)_{\Gamma}G_{32}+\left({\partial P_{23} \over \partial x_3}\right)_{\Gamma}G_{33}\right),
\end{aligned}
\label{eqn:FEMforcing2}
\end{equation}
and
\begin{equation}
\begin{aligned}
f_3^{(v)} + f_3^{\mbox{\mbox{fict}}}  & =
 \Big(\alpha (\rho-\rho_c)- \beta ((v_1^{\parallel})^2 + (v_2^{\parallel})^2+(v_3^{\parallel} \Big)^2))v_3^{\parallel} \\
& - \sigma \left({\partial \rho \over \partial x_3}\right)_{\Gamma} -\lambda \Big(v_1^{\parallel} ~{\partial  v_3^{\parallel} \over \partial x_1}+v_2^{\parallel} ~{\partial  v_3^{\parallel} \over \partial x_2}+
v_3^{\parallel} ~{\partial  v_3^{\parallel} \over \partial x_3} \\
&-v_1^{\parallel}n_3n_1 ~{\partial  v_1^{\parallel} \over \partial x_1}
-v_1^{\parallel}n_3n_2 ~{\partial  v_2^{\parallel} \over \partial x_1}
-v_1^{\parallel}n_3n_3 ~{\partial  v_3^{\parallel} \over \partial x_1}\\
&-v_2^{\parallel}n_3n_1 ~{\partial  v_1^{\parallel} \over \partial x_2}
-v_2^{\parallel}n_3n_2 ~{\partial  v_2^{\parallel} \over \partial x_2}
-v_2^{\parallel}n_3n_3 ~{\partial  v_3^{\parallel} \over \partial x_2}\\
&-v_3^{\parallel}n_3n_1 ~{\partial  v_1^{\parallel} \over \partial x_3}
-v_3^{\parallel}n_3n_2 ~{\partial  v_2^{\parallel} \over \partial x_3}
-v_3^{\parallel}n_3n_3 ~{\partial  v_3^{\parallel} \over \partial x_3}\Big)\\
&-D \left( \left({\partial P_{31} \over \partial x_1}\right)_{\Gamma} G_{11}+ \left({\partial P_{31} \over \partial x_2}\right)_{\Gamma} G_{12}+\left({\partial P_{31} \over \partial x_3}\right)_{\Gamma} G_{13}\right) \\
&-D\left(\left({\partial P_{32} \over \partial x_1}\right)_{\Gamma} G_{21}+ \left({\partial P_{32} \over \partial x_2}\right)_{\Gamma}G_{22}+\left({\partial P_{32} \over \partial x_3}\right)_{\Gamma}G_{23}\right)\\
&-D \left(\left({\partial P_{33} \over \partial x_1}\right)_{\Gamma}G_{31}+ \left({\partial P_{33} \over \partial x_2}\right)_{\Gamma}G_{32}+\left({\partial P_{33} \over \partial x_3}\right)_{\Gamma}G_{33}\right).
\end{aligned}
\label{eqn:FEMforcing3}
\end{equation}
By setting $\partial{\bf v}/\partial t$ equal to the sum of $ - \operatorname{div}_{\Gamma} {\bf J}^{(v)}$ and the ${\bf f}^{(v)}+{\bf f}^{\mbox{\mbox{fict}}}$ terms, we have
fully reproduced the curved-space version of the Toner-Tu equations.

We express the continuity equation in similar form,
\begin{equation}
{\partial \rho \over \partial t} =- \operatorname{div}_{\Gamma} {\bf J}^{(\rho)}+ f^{(\rho)} ,
\end{equation}
where ${\bf J}^{(\rho)}$ is now a vector and the scalar $f^{(\rho)}$ can be thought of as a source term. We set $\operatorname{div}_{\Gamma}{\bf J}^{(\rho)}$ equal to zero and thus define
\begin{equation}
f^{(\rho)} = - \mbox{div}_{\Gamma}  (\rho {\bf v^{\parallel}}) = -\left( {\partial (\rho v_l^{\parallel}) \over \partial x_l} -
n_ln_k {\partial (\rho v_l^{\parallel}) \over \partial x_k}\right) = - \left({\partial \rho v_1^{\parallel} \over \partial x_1}\right)_{\Gamma} - \left({\partial \rho v_2^{\parallel} \over \partial x_2}\right)_{\Gamma} - \left({\partial \rho v_3^{\parallel} \over \partial x_3}\right)_{\Gamma}.
\label{eqn:FEMcontinuity}
\end{equation}

Altogether, eqns. \ref{eqn:FEMfluxCurvedSpace}, \ref{eqn:FEMforcing1}, \ref{eqn:FEMforcing2}, \ref{eqn:FEMforcing3}, and \ref{eqn:FEMcontinuity} comprise a complete curved-surface FEM implementation of the minimal Toner-Tu equations and make possible the numerical results presented in the remainder of this work.
We also refer the interested reader to Section~\ref{SectionAppendixCOMSOL} in the Appendix for details on our implementation in the specific commercial finite element package COMSOL Multiphysics\textregistered. Our code, files, and a tutorial on their use are available at \url{https://github.com/RPGroup-PBoC/wildebeest_herds}.

\section{Parameters and Dimensionless Ratios for the Theory}
\label{Section:ParametersSection}

\subsection{Parameter Choices for Wildebeest Herds}

\begin{figure}
\centering{\includegraphics[width=5truein]{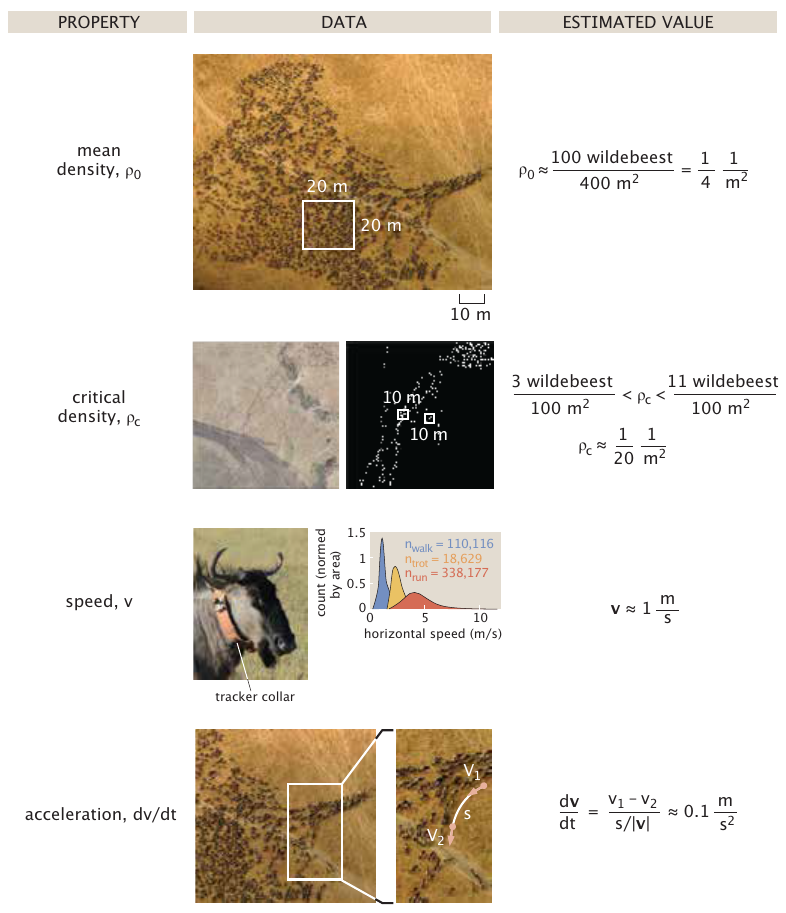}}
\caption{Estimates for key parameter values from tracking data and images of wildebeest herds. Mean density is estimated in the middle of a herd, while the critical density is estimated as in between the density of a region that does not show flocking structure (right white box) and the density of a region that does (left white box). For this study, we use a wildebeest walking speed of 1 m/s, inferred from the plot of tracking data adapted from Curtin, N {\it et al.}, (2018)~\cite{Curtin2018}. The scale of wildebeest acceleration is estimated from images of a turning herd, where a change in the vector \textbf{v} of roughly 1 m/s occurs over a time scale s/$|$\textbf{v}$|$, or roughly 10 s.  Aerial photos from drones (mean density and acceleration image) and satellites (critical density image) are from Wu, Z (2021) PhD thesis, University of Twente~\cite{Wu2021}.
\label{fig:WildebeestParameters}}
\end{figure}

To put our finite element formulation into numerical action, we must of course
adopt specific values for the parameters
$\rho_0$, $\rho_c$, $\alpha$, $\beta$, $\sigma$,  $D$, and  $\lambda$ that appear
in our governing partial differential equations.
In this study, we focus on the macroscopic length scale of animal herds, although we consider the microscopic activity of cytoskeletal proteins elsewhere~\cite{Hueschen2022c}. While our goal here is to explore the phenomenology of Toner-Tu flocks on curved surfaces in a general way, not to claim an understanding of specific animal behavior, our parameter choices are loosely inspired by migrating wildebeest herds as seen in Figure~\ref{fig:WildebeestParameters}.
By inspecting aerial photographs of wildebeest herds, we estimated an average wildebeest density of
\begin{equation}
\rho_0 = 0.25~\mbox{m}^{-2}.
\end{equation}
For the critical density above which coordinated herding behavior occurs, we make the estimate
\begin{equation}
\rho_{c} = 0.05~\mbox{m}^{-2}, 
\end{equation}
based roughly on the observation that
for densities much higher than this, the photographed wildebeest have an organized
herd structure.  We note that it would be very interesting to carefully measure these parameters
in the context of herds of wildebeest with a Toner-Tu framework in mind.

The remainder of the parameters are determined in the spirit of exploring how the different terms
compete to alter the density and velocity fields.
Thus, parameters are chosen in order to make all the terms comparable in magnitude.
We provide here the actual values used in our finite element calculations, fully cognizant
that our parameter choices are at best approximate.
We begin by estimating the magnitude of $\partial v/\partial t$.
We picture wildebeests circling a small hill in the landscape, moving with a characteristic speed of 1~m/s and taking 20 s to change direction completely. This scenario implies
\begin{equation}
{\partial v \over \partial t} \approx {1~\mbox{m/s} - (-1~\mbox{m/s}) \over 20~\mbox{s}} \approx 0.1~\mbox{m/s}^2.
\end{equation}
Using the strategy of balancing the magnitudes of the different terms, we estimate the coefficient for the preferred speed terms by considering that
\begin{equation}
0.1~\mbox{m/s}^2 = \alpha (\rho_0-\rho_c) v \approx \alpha \times (0.25~\mbox{m}^{-2}-0.05~\mbox{m}^{-2}) \times 1~\mbox{m/s} ,
\end{equation}
which leads us to adopt
\begin{equation}
\alpha = 0.5~\mbox{m$^2$/s}.
\end{equation}
For the parameter $\beta$, we choose a value that sets the correct mean field speed, $v_{\mbox{\footnotesize{pref}}}$, for
wildebeests. Thus, we adopt
\begin{equation}
\beta=\frac{\alpha\left(\rho_0-\rho_{c}\right)}{v_{\mbox{\footnotesize{pref}}}^{2}} \approx \frac{\left(0.5 ~\mbox{m}^2/\mbox{s} \right) \left(0.2 ~\mbox{m}^{-2}\right)}{(1 \mathrm{~m} / \mathrm{s})^{2}} = 0.1 \mathrm{~s} / \mathrm{m}^{2}.
\end{equation}
We can also check independently that the magnitude of the $\beta$ term is of order 0.1 $\mbox{m}/\mbox{s}^2$. Indeed,
\begin{equation}
\beta  v^3 \approx 0.1 ~{\mbox{s} / \mbox{m}^2}  \times  (1~\mbox{m/s})^3 =0.1~\mbox{m/s}^2.   
\end{equation}
To find the parameter $\sigma$, the coefficient of the pressure term, we consider the
gradients in density seen at the edge of a wildebeest herd and estimate that $\rho$ drops from $\rho_0 = 0.25 ~\mbox{m}^{-2}$ to 0 $\mbox{m}^{-2}$ over 25~m.  Using these numbers implies
\begin{equation}
0.1~ \mbox{m/s}^2 =\sigma {\partial \rho \over \partial x}  \approx \sigma {0.25~\mbox{m}^{-2}  \over 25~\mbox{m}}
\end{equation}
which leads us to adopt
\begin{equation}
\sigma = 10~\mbox{m}^4/\mbox{s}^2.
\end{equation}
We can now apply this thinking to make an estimate of the neighbor coupling coefficient $D$ by using
the equality
\begin{equation}
0.1~\mbox{m/s}^2 = D {\partial^2 v \over \partial x^2}  \approx D{1~\mbox{m/s}   \over (10~\mbox{m})^2} 
\end{equation}
which leads to the coefficient choice
\begin{equation}
D = 10~ \mbox{m}^2/\mbox{s}.
\end{equation}
To estimate the magnitude of $\lambda$, we imagine
 that the wildebeest will come to a full stop from a speed of 1~m/s over a distance of roughly 10 m, 
 permitting us to make the correspondence
\begin{equation}
0.1~\mbox{m/s}^2 = \lambda v {\partial v \over \partial x} \approx \lambda \times 1~ \mbox{m/s}\times {1~ \mbox{m/s}\over 10~ \mbox{m}}
\end{equation}
which implies that
\begin{equation}
\lambda = 1.
\end{equation}
This estimated set of parameters provides us with a complete description of the Toner-Tu herd in the minimal model
we seek to explore. Throughout the remainder of the paper, we consistently use these parameter values.  In later sections, in some cases
we will go beyond this idealized parameter set to perform parameter sweeps that permit
different regimes of behavior to emerge. Unless noted otherwise, time-dependent finite element method simulations were initialized with a uniform density field $\rho = \rho_{0}$ and a disordered velocity field, with velocity orientations drawn randomly from a uniform distribution of angles between 0 and 2$\pi$ and with velocity magnitude $v(0)$ = 1 m/s, our estimated characteristic wildebeest speed. We note again that these parameters were chosen to highlight competition
between the different terms in the dynamics; they are far from the final word for describing real wildebeest herds.

\subsection{Dimensionless Representation of the Theory}

We next recast the Toner-Tu equations in dimensionless form,
rendering the meaning and magnitude of the terms more transparent. For the positional coordinate,
we take
$x^* = x/L$,
where $L$ is a characteristic length scale in the problem. In this case, we conceptualize $L$ as the length scale of the wildebeest herd. We note that in the problems considered here, the scale of the herd is the same as the scale of the surface landscape on which the herd moves.
Similarly, we use a characteristic velocity scale $U$ to define
the dimensionless velocity as 
$v^*= v/U$.  We note that $U$ is conceptually related to $v_{\mbox{\footnotesize{pref}}}$, but $v_{\mbox{\footnotesize{pref}}}=\sqrt{\alpha(\rho-\rho_c)/\beta}$ features a hidden dependence on the
density that would complicate our rescaling.  
Similarly, we can define a dimensionless density $\rho^* =\rho /\rho_c$, using the critical density $\rho_c$ as our scaling variable.
Lastly, in light of the definitions above, we can determine a time scale $L/U$, the time it takes for a given wildebeest to cross the entire surface landscape. Thus, we define
$t^*=t/(L/U)$.

Using the various definitions given above, we can now rewrite the Toner-Tu equations using
the dimensionless versions of $t$, $x$, $\rho$ and ${\bf v}$ as
\begin{equation}
{U^2 \over L} {\partial v_i^* \over \partial t^*} =
\alpha \rho_c U (\rho^*-1) v_i^*-\beta  U^3 |{\bf v}^*|^2 v_i^* -{\sigma \rho_c \over L}{\partial  \rho^* \over \partial x_i^*}  + {D  U \over L^2} \nabla_*^2 v_i^*- {\lambda  U^2 \over L}  v_j^* {\partial  v_i^* \over \partial x_j^*}.
\end{equation}
Dividing everything by $U^2/L$ results
in five dimensionless parameters whose magnitudes provide a sense of the relative contributions of the different terms, within the full dynamical equations of the form
\begin{equation}
{\partial v_i^* \over \partial t^*} =
{\alpha \rho_c L \over U} (\rho^*-1) v_i^*-\beta L U |{\bf v}^*|^2 v_i^*-{\sigma \rho_c \over U^2} {\partial  \rho^* \over \partial x_i^*}  + {D   \over UL} \nabla_*^2 v_i^*  - \lambda v_j^* {\partial  v_i^* \over \partial x_j^*}.
\label{eqn:TonerTuDimensionless1}
\end{equation}
Using definitions of dimensionless variables
given above, 
we find that the generalized continuity equation takes the form
\begin{equation}
{\partial \rho^* \over \partial t^*} =- {\partial (\rho^* v_i^*)\over \partial x_i^*}.
\end{equation}

Next, we briefly turn to interpreting the dimensionless ratios that appear when casting
the minimal Toner-Tu theory in dimensionless form as exhibited in
eqn.~\ref{eqn:TonerTuDimensionless1}. 
The two components of the preferred speed term have dimensionless parameters given by
\begin{equation}
\mbox{increase to preferred speed parameter}= {\alpha \rho_c L \over U} = {L/U \over 1/(\alpha \rho_{c})} = {\mbox{time for wildebeest to cross herd} \over
	\mbox{time for speed to increase to $v_{\mbox{\footnotesize{pref}}}$ } }
\end{equation}
and
\begin{equation}
\mbox{decrease to preferred speed parameter}= \beta L U = {L/U \over 1/(\beta U^2)} = {\mbox{time for wildebeest to cross herd} \over
	\mbox{time for speed to decrease to $v_{\mbox{\footnotesize{pref}}}$}}.
\end{equation}
Each of these terms provides intuition about how quickly the wildebeest will return to their
steady-state speed given some perturbation that disturbs them from that value. Recall that $L/U$ (the ``time for wildebeest to cross herd") is roughly the time it takes for a wildebeest to move a distance equal to the size of the herd, or equivalently, to move across the surface landscape. Put in other words, $L/U$ is the time for density advection across the herd.
The pressure term can be rewritten as
\begin{equation}
\mbox{pressure parameter}= {\rho_c \over U^2/\sigma}={\mbox{critical density} \over
	\mbox{typical density excursion away from $\rho_{0}$ }}.
\end{equation}
Increasing $\sigma$ decreases density variance; in other words, densities that emerge are within a more narrow range around the mean density $\rho_{0}$.
The neighbor coupling term that carries out democratic velocity smoothing has the dimensionless prefactor
\begin{equation}
\mbox{neighbor coupling parameter}={D\over U L} = {L/U \over L^2/D} ={\mbox{time for wildebeest to cross herd} \over
	\mbox{time for velocity to diffuse across herd}},
\end{equation}
analogous to the P\'{e}clet number.  The last term can be written as
\begin{equation}
\mbox{advection parameter}={\lambda L/U  \over L/U}={\mbox{time for velocity advection across herd} \over
	\mbox{time for wildebeest to cross herd (density advection)}}.
\end{equation}

These dimensionless ratios give a sense of
how large a contribution a given term will make to the incremental update to
${\bf v}(t)$.    
In a sense, their values capture the relative importance
of each term, serving roles analogous to the Reynolds number in thinking
about the Navier-Stokes equations and the P\'{e}clet number in the context of
coupled diffusion-advection problems.

\section{Solving Toner-Tu on Highly Symmetric Geometries}
\label{Section:AnalyticSolutions}

In this section, we use the finite element implementation of the curved-space minimal Toner-Tu model to illustrate herding in symmetric geometries in anticipation of the fully general case. Our goals are two-fold: to explore the behavior of Toner-Tu herds in classic geometries but also, importantly, to validate our finite element implementation by comparison to analytic solutions. This validation is particularly important in the context of systems as complex as the partial differential equations considered here, which are already subtle in flat space and more demanding yet in their curved-space form.  We begin with the planar geometry of a two-dimensional (2D) channel before moving to a 2D channel with an embedded scattering obstacle. Then, we turn to collective herding motions on cylindrical and spherical surfaces, as those two cases do admit analytic solutions that can be directly compared to the curved-space numerical results.

\subsection{Solving the Toner-Tu Equations in the Plane}

One of the classic case studies for  traditional fluid mechanics is channel flow, a problem
dating back at least to the 19th century.  Here, we consider several versions of
the channel flow problem in the context of active herding agents, with the recognition that
herding animals are sometimes faced with passing through canyons or gorges. 
In the Disney movie {\it The Lion King}, Hollywood filmmakers gave their
own version of herding behavior in a scene in which the young lion Simba
is trapped in a gorge of stampeding wildebeest as a result of the 
machinations of his evil uncle Scar.  In this section, in a playful frame of mind,
we use the minimal Toner-Tu theory in conjunction with the finite element method
to explore a herd of stampeding wildebeest in a gorge. We first consider a 2D
channel, yielding the Toner-Tu version of pipe flow. This problem has been explored experimentally in
work on annular channels~\cite{ Wioland2016, Souslov2017, Wu2017, Morin2018}. We follow this by an analysis with an obstacle in
the middle of the channel \cite{Zhang2020a} to show the generality and flexibility of the finite element implementation in action.

\begin{figure}
	\centering{\includegraphics[width=5.5truein]{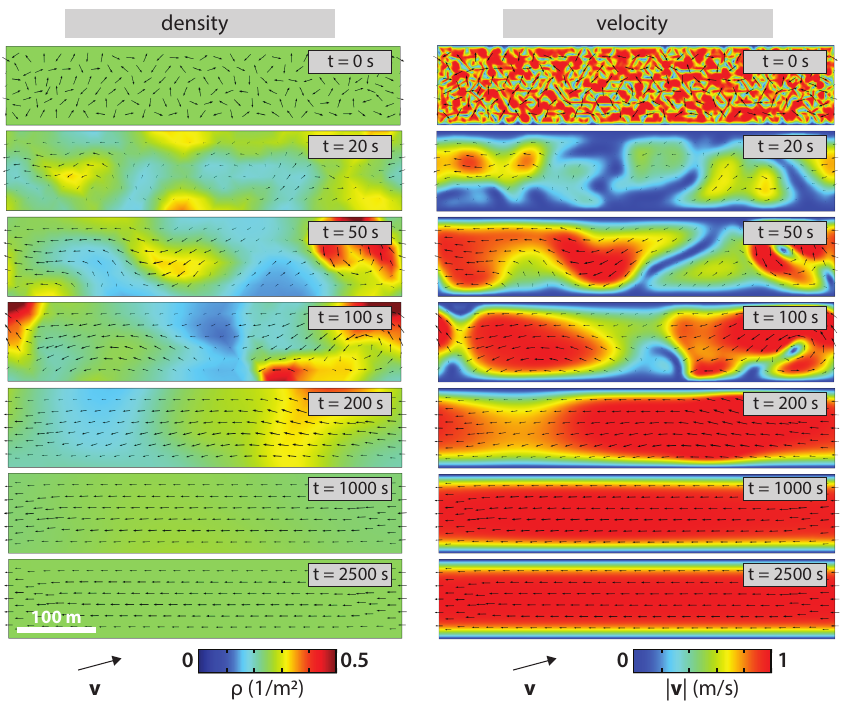}}
	\caption{Toner-Tu herding in a 2D channel with periodic boundary conditions at its ends. The left panel shows the time evolution
		of the density field (colors) with the velocities superimposed as arrows.  The right panel reports on the same simulation but uses a color map
		to show the speed.  In the long-time limit, the density is uniform and the speed is constant throughout
		the channel except for a boundary layer near the walls. See also Video 1.
		Here and in all subsequent figures, parameters used are those inferred in  Section~\ref{Section:ParametersSection} unless explicitly stated otherwise.
		\label{fig:ChannelFlow}}
\end{figure}

Solving the problem of channel flow is a rite of passage in the study of the
hydrodynamics of Newtonian fluids.  However, for active agents, this simple geometry already reveals interesting phenomenology, as shown in Figure~\ref{fig:ChannelFlow} and Video 1.
As illustrated in the figure, we begin with a uniform density and a randomly
oriented velocity field in a 2D channel with periodic boundary conditions at its ends. We then use our finite element implementation of
the minimal Toner-Tu theory to explore the transient and steady-state dynamics.
Despite starting from an initially disordered velocity field,
 in the long-time limit the system finds a steady state with uniform
density $\rho_0$ and speeds that are consistent with the preferred speed term, here equivalent to
$|{\bf v}|=\sqrt{\alpha \left(\rho_0-\rho_{c}\right)/\beta}$.  Near the edges of the channel,
because of the no-slip boundary condition, there is a boundary layer
interpolating between the optimal speed at the middle of the channel
and the zero velocity at the walls.   We note that it is not at all
clear that no-slip boundary conditions are the most reasonable
biological choice either for animal herds in a gorge or for cytoskeletal filaments in
channels, but solving this problem provides a 
case study for exploring the intuition behind the Toner-Tu theory
and our numerical implementation of it.

\begin{figure}
\centering{\includegraphics[width=4truein]{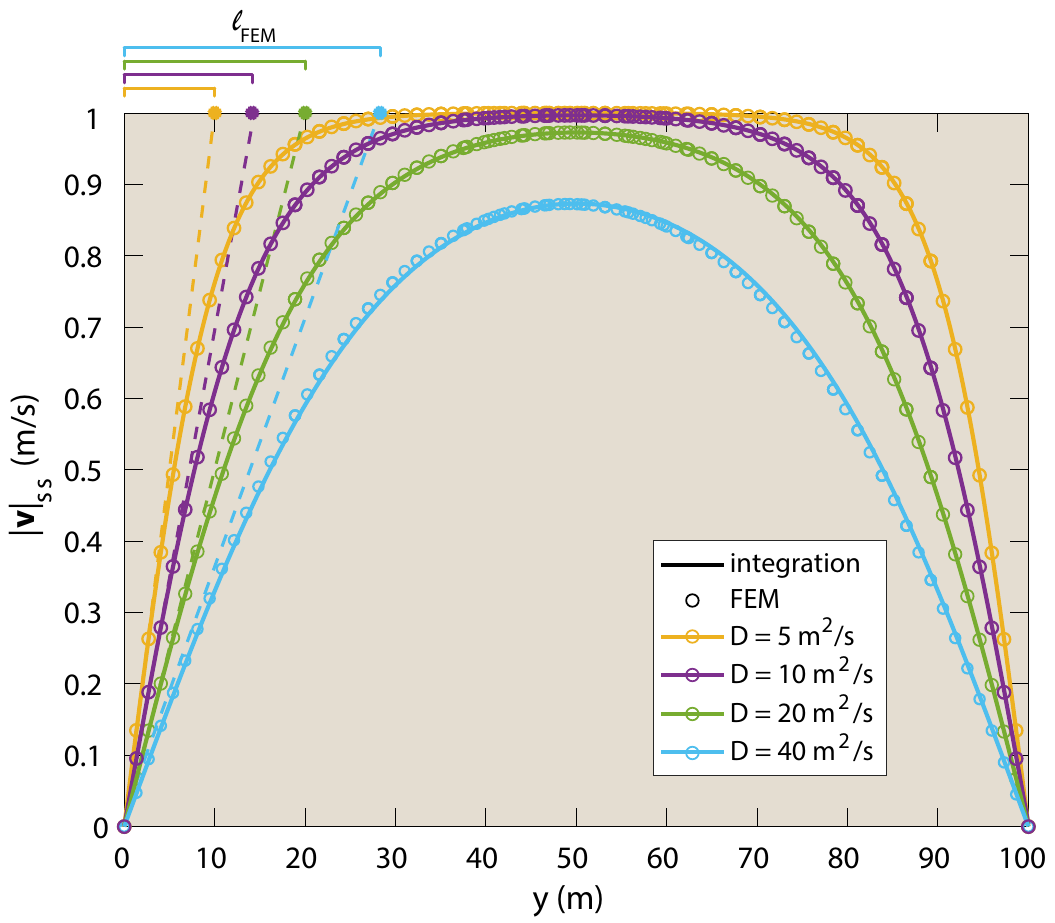}}
\caption{Steady-state velocity for active agents during channel flow as a function of position y, across the channel. Solutions were obtained
both by the time-dependent finite element method (circles) and by numerically integrating the steady-state Toner-Tu equations (solid lines)
for the highly symmetric channel geometry.  Different choices of the neighbor coupling coefficient $D$ result in boundary layers of different thickness.  The width $\ell_{\mbox{FEM}}$ of the boundary layer
is estimated by computing the initial slope of the velocity vs position curves and extrapolating out (dashed lines) to the preferred speed ($v_{\mbox{\footnotesize{pref}}} = 1~$m/s), as shown in the top left of the figure. In Figure \ref{fig:ChannelBoundaryLayer},  $\ell_{\mbox{FEM}}$ is compared to analytical calculations of the boundary layer thickness.
\label{fig:ChannelFEMvsShooting}}
\end{figure}

Tuning the neighbor coupling coefficient $D$ results in different steady-state velocity profiles, $v_{ss}(y)$, as shown in Figure \ref{fig:ChannelFEMvsShooting}. For comparison to these finite element method findings, we now consider several analytical calculations. In steady state, the velocity field in the minimal
Toner-Tu model for
this channel geometry is described by
\begin{equation}
D{d^2 v_x(y) \over dy^2}+ \Big(\alpha (\rho_0-\rho_c) -\beta v_x^2(y) \Big) v_x(y)=0.
\label{eqn:ChannelSteadyState}
\end{equation}
We can solve this equation by numerical integration using ``shooting" methods. 
As shown in Figure~\ref{fig:ChannelFEMvsShooting},
the time-dependent finite element method and the
shooting method solutions to eqn. \ref{eqn:ChannelSteadyState} give essentially indistinguishable results.
This correspondence provides a validation of the FEM approach.

\begin{figure}
 	\centering{\includegraphics[width=3.5truein]{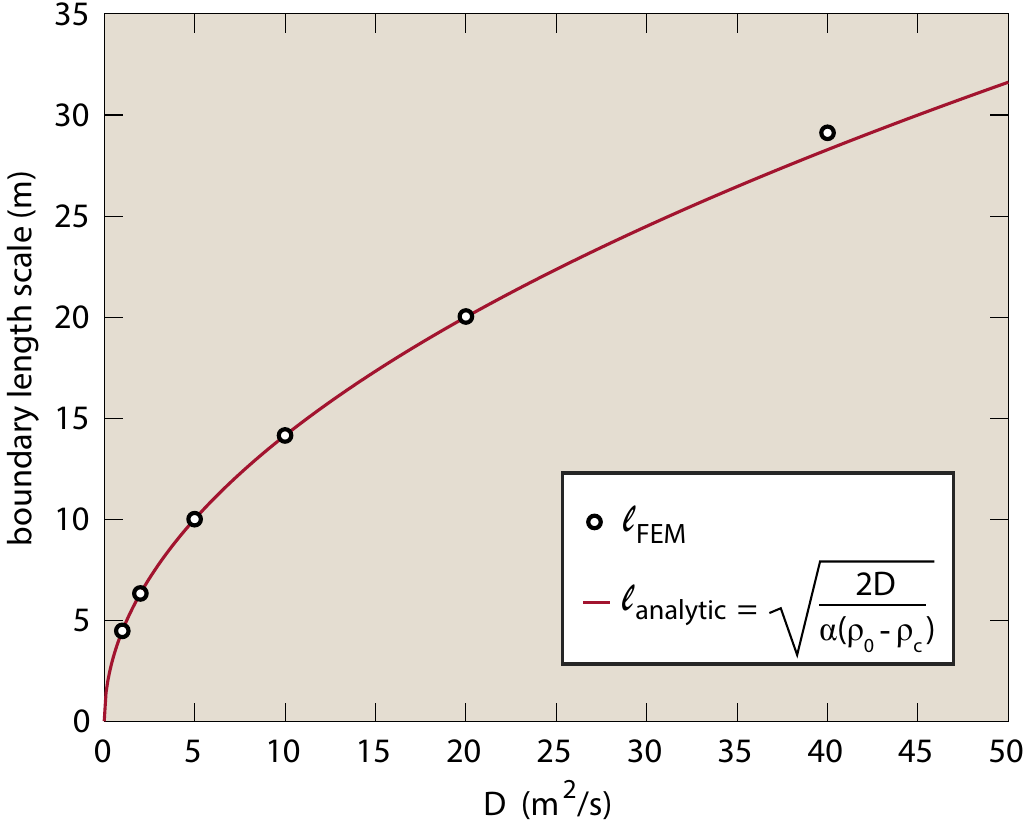}}
 	\caption{Boundary layer thickness as a function of the neighbor coupling coefficient, $D$.  A numerical result for the thickness of
 		the boundary layer, $\ell_{\mbox{FEM}}$, was estimated by computing $dv/dy$ at the wall and using the condition
 		$(dv/dy) \ell_{\mbox{FEM}} =v_{\mbox{\footnotesize{pref}}}$, where 
 		$v_{\mbox{\footnotesize{pref}}}=\sqrt{\alpha \left(\rho_0-\rho_{c}\right)/\beta}$. An analytical result for the thickness of the boundary layer, $\ell_{\mbox{analytic}}$, was calculated in eqns. \ref{eqn:HalfSpace1}-\ref{eqn:HalfSpace4}.
 		\label{fig:ChannelBoundaryLayer}}
 \end{figure}

For a deeper understanding of the channel problem, we focus on the velocity boundary layers clearly noticeable in Figure \ref{fig:ChannelFEMvsShooting}. For sufficiently small values of
the neighbor coupling coefficient $D$, 
the preferred speed term keeps the steady-state speed at $v_{\mbox{\footnotesize{pref}}}$ = 1 m/s
throughout the channel except for a narrow band near the walls,
where the no-slip condition gives rise to a boundary layer. 
In Figure \ref{fig:ChannelFEMvsShooting}, we show how the width of the boundary layer in the finite element method solution,  $\ell_{\mbox{FEM}}$, can be estimated using the condition
\begin{equation}
{dv_x(y) \over dy} \ell_{\mbox{FEM}} \bigg\rvert_{y=0} = v_{\mbox{\footnotesize{pref}}},
\end{equation}
where $v_{\mbox{\footnotesize{pref}}}=\sqrt{\alpha(\rho_0-\rho_c)/\beta}$ is the preferred speed.  
We explore the scaling of boundary layer thickness numerically for different values of the neighbor coupling coefficient $D$ using
the finite element implementation, and we report these numerically-determined boundary layer widths, $\ell_{\mbox{FEM}}$, in Figure~\ref{fig:ChannelBoundaryLayer}.

To interpret the length scale of the boundary layer 
and its scaling with $D$, we can perform an estimate by introducing the dimensionless variables 
\begin{equation}
\bar{y}= {y \over   \sqrt{{D \over \alpha (\rho_0-\rho_c)}}},
\label{eqn:LengthScale}
\end{equation}
and
\begin{equation}
\bar{v}= {v_x \over   \sqrt{{ \alpha (\rho_0-\rho_c) \over \beta}}}.
\label{eqn:VelocityScale}
\end{equation}
With these dimensionless variables in hand, we can rewrite
the steady-state equation for velocity in the channel (eqn.~\ref{eqn:ChannelSteadyState}) in dimensionless form
as 
\begin{equation}
{d^2 \bar{v} \over d\bar{y}^2}+(1 -\bar{v}^2)\bar{v}=0.
\label{eqn:ChannelSteadyStateDimensionless}
\end{equation}
The act of rendering the equation in dimensionless form immediately presents us
with the length scale 
\begin{equation}
\ell = \sqrt{D \over \alpha (\rho_0-\rho_c)},
\label{eqn:BoundaryLengthscaleDimAnalysis}
\end{equation}
which gives us a sense of the thickness of the boundary layer found by numerical methods in Figure~\ref{fig:ChannelFEMvsShooting}.

%
This scaling relation can be understood more rigorously by considering
the problem of the velocity $v_x(y)$ of active agents in a half space  for $0 \le y <  \infty$.
In this case, we have $v_x(0)=0$ as a no-slip boundary condition
and $v_x(\infty)=\sqrt{ \alpha (\rho_0-\rho_c)/ \beta}$, implying  that far from the wall, the speed achieves the Toner-Tu preferred speed value.
In this case, eqn.~\ref{eqn:ChannelSteadyStateDimensionless} has the analytic
solution $\bar{v}(\bar{y})=\mbox{tanh}~(\bar{y}/\sqrt{2})$, which can be written in
dimensionful form as
\begin{equation}
v_x(y)= \sqrt{ \alpha (\rho_0-\rho_c) \over \beta}~\mbox{tanh}~\sqrt{{ \alpha (\rho_0-\rho_c) \over 2D}} y.
\label{eqn:HalfSpace1}
\end{equation}
For sufficiently small $y$ we can approximate this by Taylor expanding the
tanh function to first order in $y$, resulting in
\begin{equation}
v_x(y) \approx {\alpha (\rho_0-\rho_c) \over \sqrt{2 \beta D}} y
\end{equation}
which implies in turn that the thickness of the  boundary layer can be estimated using
\begin{equation}
 \frac{d v_{x}}{d y} \bigg\rvert_{y=0} \ell_{\mbox{analytic}}=\frac{\alpha\left(\rho_0-\rho_{c}\right)}{\sqrt{2 \beta D}} \ell_{\mbox{analytic}}=\sqrt{\frac{\alpha \left(\rho_0-\rho_{c}\right)}{\beta}}.
\end{equation}
Solving for the length scale of the boundary layer by projecting
the straight line with the correct slope at $y=0$ out to the saturation speed, as done in
Figure~\ref{fig:ChannelFEMvsShooting} to compute the analogous $\ell_{\mbox{FEM}}$, leads to
\begin{equation}
\ell_{\mbox{analytic}}=\sqrt{\frac{2 D}{\alpha \left(\rho_0-\rho_{c}\right)}},
\label{eqn:HalfSpace4}
\end{equation}
with the factor of $\sqrt{2}$ arising naturally from
the analytic solution for the half space. This analytical calculation for the boundary layer length 
$\ell_{\mbox{analytic}}$ jibes
exactly with the numerical result $\ell_{\mbox{FEM}}$, as seen in Figure~\ref{fig:ChannelBoundaryLayer}, until $D$ reaches values large enough that the half-space assumption of  $v_x(\infty)=\sqrt{ \alpha (\rho_0-\rho_c)/ \beta}$ becomes invalid.

\begin{figure}
\centering{\includegraphics[width=5.5truein]{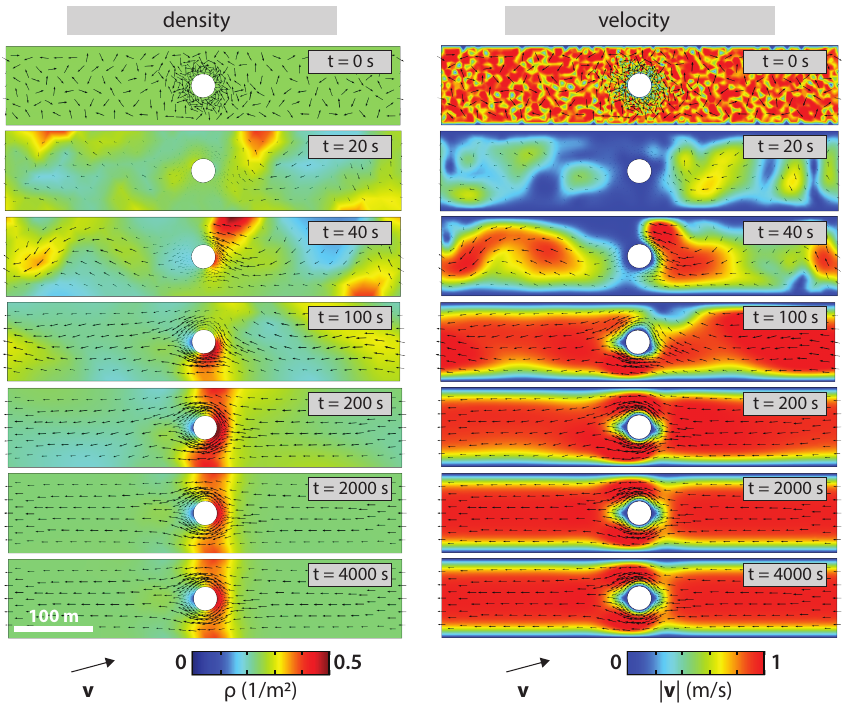}}
\caption{Solution of the Toner-Tu equations for a herd moving through a channel with a small circular obstacle (15 m obstacle radius; 100 m channel width).  The left panel shows the time evolution
of the density field (colors) with the velocities superimposed as arrows, from an initial condition of
uniform density and a disordered velocity field.  The right panel uses a color map
to show the speed. For this choice of obstacle size and neighbor coupling coefficient $D$, a steady-state flow is
achieved in which the active agents can reach their preferred speed even in the passage
between the obstacle and the wall. See also Video 1.
\label{fig:ChannelFlowObstacle}}
\end{figure}

\begin{figure}
\centering{\includegraphics[width=5.5truein]{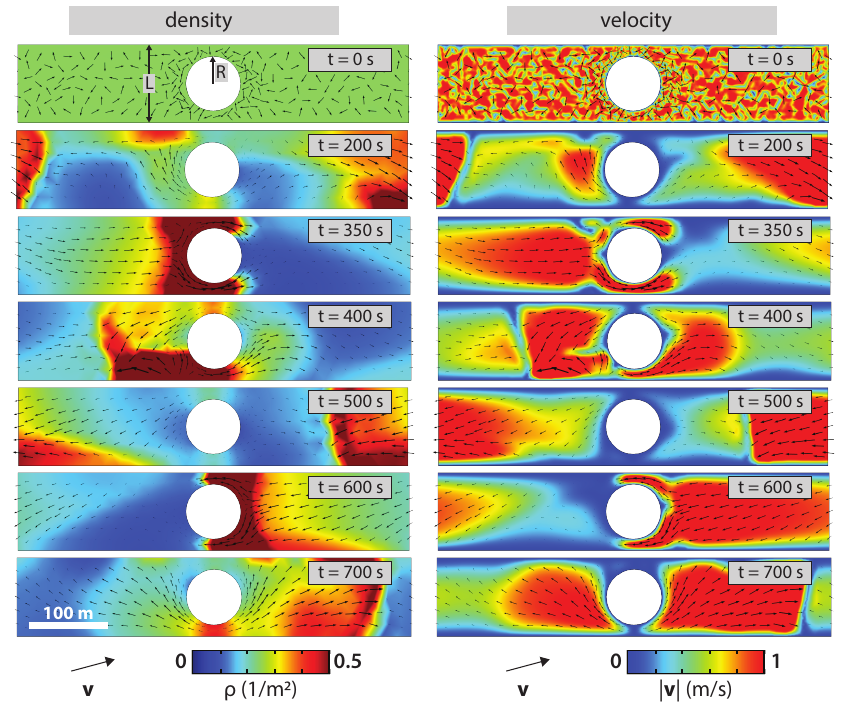}}
\caption{Solution of the Toner-Tu equations for a herd moving through a channel with a large circular obstacle (30 m obstacle radius; 100 m channel width).  The left panel shows the time evolution
of the density field (colors) with the velocities superimposed as arrows.  The right panel uses a color map
to show the speed.   For this choice of obstacle size and neighbor coupling coefficient $D$, no steady
state motion is achieved. Most of the herd is reflected off the obstacle at each attempted passage, resulting in repeated reversals of direction. See also Video 1.
\label{fig:ChannelFlowObstacle2}}
\end{figure}

The problem of Toner-Tu channel flow for wildebeest becomes richer when we introduce a circular obstacle into the channel,
as shown in Figure~\ref{fig:ChannelFlowObstacle} and Video 1.  In the classic example of this problem for
a Newtonian fluid, the fluid motion will speed up near the obstacle because of the narrowing of the channel.  However, in the case of a Toner-Tu collective,
the preferred speed term attempts to maintain all agents at a fixed speed.
Thus, if a large obstacle narrows the channel sufficiently, as shown in  Figure~\ref{fig:ChannelFlowObstacle2} and Video 1,
most of the herd will reflect off the obstacle rather than squeezing through at a higher velocity. Interestingly, this oscillatory direction reversal at an obstacle was observed in recent experimental and theoretical work, in which rectangular or triangular objects partially blocked the path of an active colloidal fluid confined in a ring-shaped track \cite{Zhang2020a}.

To better understand the origins of these reversals of motion, we contrast the Toner-Tu
theory with our expectations for Newtonian fluids.  Specifically, we note the crucial role of the preferred speed term, which penalizes any active agents with velocity magnitudes other than
 $v=\sqrt{\alpha \left(\rho-\rho_{c}\right)/\beta}$.    
We estimate the critical obstacle radius $R$ for a channel of width $L$ by imagining that when the thickness of the passage
that the active agents can pass through (i.e. $L/2 -R$)  is comparable
to the boundary layer thickness, none of the active agents get to move
at the optimal speed and as a result will reflect off the obstacle.
That condition can be written as 
\begin{equation}
\frac{L}{2}-R=2 \sqrt{\frac{D}{\alpha \left(\rho_0-\rho_{c}\right)}},
\label{eqn:ChannelObstacle}
\end{equation}
where the factor of two on the right side captures the idea that
if the gap is wide enough, there will be two boundary layers, one at the
obstacle and one at the wall.  When the width of these {\it two} boundary layers
adds up to the width of the gap, we estimate that
reversals will occur.  Note that as written, eqn.~\ref{eqn:ChannelObstacle} features
the length scale $\ell$ that arose from recasting the steady-state Toner-Tu equations in
dimensionless form as defined in eqns.~\ref{eqn:LengthScale} and ~\ref{eqn:VelocityScale}, not the length scale $\ell_{\mbox{analytic}}$ from our analytic solution for the half space.

\begin{figure}
\centering{\includegraphics[width=3.5truein]{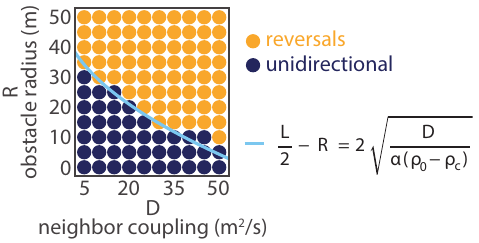}}
\caption{Phase diagram of herding behavior in a channel with an obstacle.  Obstacle radius, R, and neighbor coupling coefficient, D, were systematically varied. Emergent herd dynamics were either unidirectional, reaching a steady-state as in Figure~\ref{fig:ChannelFlowObstacle}, or displayed motion reversals at the obstacle indefinitely, as in Figure~\ref{fig:ChannelFlowObstacle2}. The transition between these two behaviors appears to occur when the width of the passage between obstacle and channel wall, $L/2 - R$, is comparable to twice the boundary layer length scale $\ell$ obtained by dimensional analysis in eqn. \ref{eqn:BoundaryLengthscaleDimAnalysis}.
\label{fig:ChannelFlowObstacle3}}
\end{figure}

As seen in Figure~\ref{fig:ChannelFlowObstacle3}, we have explored
the phenomenology of the motions in the channel with an obstacle by sweeping through the parameter
space of obstacle size $R$ and neighbor coupling coefficient $D$.  Depending
upon the choices of these parameters as already seen in Figures~\ref{fig:ChannelFlowObstacle}
and ~\ref{fig:ChannelFlowObstacle2}, we find either steady unidirectional motion or dynamical reversals, which continue indefinitely.
Interestingly, the boundary in this phase portrait is in very good accord with
the estimate of eqn.~\ref{eqn:ChannelObstacle}. 
In light of the apparent success of these case studies on the minimal Toner-Tu model using a finite element
implementation for planar geometries, we turn to the application of the minimal Toner-Tu model and its
finite element implementation 
to curved surfaces. 

\subsection{Formulating the Toner-Tu Equations for Parameterized Surfaces}
\label{Section:TonerTuForParameterizedSurfaces}

We now move to apply and test our general surface formulation (Section \ref{Section:FlockingTheory}) and finite element implementation (Section \ref{Section:FiniteElement}) of the Toner-Tu equations on curved surfaces. We begin with two highly symmetric surfaces, the cylinder and the sphere, for which it is tractable to derive analytical solutions using intrinsic parameterizations. In the remainder of Section \ref{Section:AnalyticSolutions}, we will analytically determine dynamical Toner-Tu solutions for velocity on the cylinder and steady-state Toner-Tu solutions for density and velocity on the sphere. These analytical solutions are of great interest to us as testbeds for
our curved-space finite element formulation of the minimal Toner-Tu theory. We will compare our analytical calculations to results obtained using the finite element approach, with the ultimate goal of validating its performance.

First, we must formulate the minimal flocking theory in the language of intrinsic differential geometry. We follow the beautiful analytical study by Shankar {\it et al.}~\cite{Shankar2017}  of the Toner-Tu equations on
highly symmetric curved surfaces, in which steady-state solutions for the sphere were considered.  
Following Shankar {\it et al.}~\cite{Shankar2017}, we need to write down the fully covariant form of
the dynamical Toner-Tu equations.  
First, the continuity equation can be written as
\begin{equation}
\partial_{t} \rho+\nabla_{\mu} (\rho v^{\mu})=0.
\end{equation}
This formulation  invokes the covariant derivative
\begin{equation}
\nabla_{\mu} v^{\nu}=\partial_{\mu} v^{\nu}+\Gamma_{\alpha \mu}^{\nu} v^{\alpha}
\end{equation}
where $\Gamma_{\alpha \mu}^{\nu}$ are the  Christoffel symbols.  All of this extra machinery
and corresponding notation comes down to the subtlety of getting our derivatives
right in the curved-space setting.  
Then, the dynamical equation for ${\bf v}$ itself is given by
\begin{equation}
{\partial v^{\mu} \over \partial t}=\left[\alpha \left(\rho-\rho_{c}\right)-\beta g_{a b} v^{a} v^{b}\right] v^{\mu}-\sigma \nabla^{\mu} \rho+ D \Delta v^{\mu}-\lambda v^{\nu} \nabla_{\nu} v^{\mu}
\label{eqn:FullTonerTuCurved}
\end{equation}
where we introduce the symbol $\Delta$ to represent the intrinsic curved-space version of
the Laplacian operator.
We now put these equations into concrete play in the context of
cylinders and spheres, with the ambition of directly comparing analytical results and finite element
solutions for the same problems.

\subsection{Solving the Toner-Tu Equations on a Cylindrical Surface}
\label{SectionCylinderAnalytical}

To test our numerical implementation of the Toner-Tu equations in the finite element setting
for curved surfaces, we begin with the highly symmetrical case of a cylindrical surface.  The geometry of the cylindrical surface is parameterized using
${\bf r}(\theta, z)=(R \cos \theta, R \sin \theta, z)$.
Using this parameterization, we seek to analyze the steady-state and the dynamics of ${\bf v}$ for a simplified version of eqn.~\ref{eqn:FullTonerTuCurved},
\begin{equation}
{\partial v^{\mu} \over \partial t}=\left[\alpha \left(\rho-\rho_{c}\right)-\beta g_{a b} v^{a} v^{b}\right] v^{\mu}-\sigma \nabla^{\mu} \rho-\lambda v^{\nu} \nabla_{\nu} v^{\mu},
\end{equation}
 in which the neighbor coupling term with coefficient $D$ is neglected.
 In Appendix Section \ref{SectionAppendixCylinderAnalytical}, we first present a complete derivation of the steady-state solution of this equation for the cylinder, which leads to the result
\begin{equation}
\left|\mathbf{v}_{s s}\right|^{2}={\alpha \left(\rho_0-\rho_{c}\right) \over \beta },
\end{equation}
where ${\bf v}_{ss}$ is the steady-state velocity field, 
whose only non-zero component is $v^{\theta}$.
This solution represents a simple circumferential flow around the cylinder
coupled to a uniform density field. 

We then derive the dynamics of a relaxation to the steady-state, from an initial condition with the same symmetry
of circumferential flow and uniform density, but with an arbitrary initial speed. 
Given an initial magnitude $v(0)$
and a uniform density $\rho$, the solution for the time dependent relaxation is given by 
\begin{equation}
\left|\mathbf{v}\right|^{2}=\frac{\frac{\alpha \left(\rho_0-\rho_{c}\right)}{\beta}}{1-\left(1-\frac{\alpha \left(\rho_0-\rho_{c}\right)}
{\beta ~v(0)^{2}}\right) e^{-2 \alpha \left(\rho_0-\rho_{c}\right) t}}.
\end{equation}

\begin{figure}
\centering{\includegraphics[width=5.5truein]{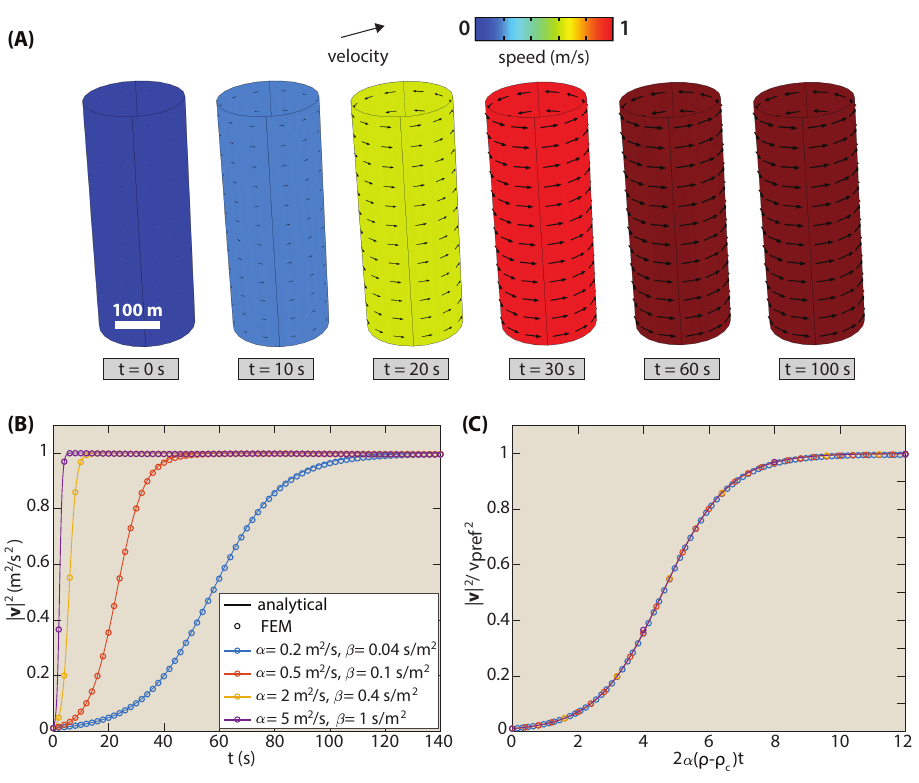}}
\caption{Dynamics of Toner-Tu flow on a cylinder. (A) The initial condition is a uniform circumferential
flow field with a very small initial speed.  Over time, the speed relaxes to the steady-state
value $|{\bf v}_{ss}|=\sqrt{\alpha \left(\rho_0-\rho_{c}\right) /\beta}$.  The time scale for this relaxation is
$\tau=1/(2 \alpha \left(\rho_0-\rho_{c}\right) )$.  (B) Comparison of analytic and numerical (COMSOL) results for dynamics of $|{\bf v}|^2$ for different choices of parameters.  The asymptotic values of the curves correspond
to the steady-state solution worked out earlier in this section.  The ratio $\alpha/\beta$ is fixed such that the steady-state velocity of 1 m/s is the same for all four cases, and $|{\bf v}|(0)$ = 0.1 m/s.   (C) Data collapse in which $|{\bf v}|^2$ is plotted against dimensionless
time, revealing that all the curves are dictated by the same underlying time scale, set by $\alpha \left(\rho_0-\rho_{c}\right)$.
\label{fig:DynamicsCylinder}}
\end{figure}
In Figure~\ref{fig:DynamicsCylinder}, we compare the dynamics predicted here analytically to the numerical solution of our curved-surface formulation using the finite element method.
As with many theoretical analyses, one of the most important aspects of our solutions is understanding
how the phenomenology depends upon parameters.  Here we see how the parameter $\alpha$ controls
the saturation value of $|{\bf v}_{ss}|^2$  as well as the time scale ($\tau=1/2\alpha  \left(\rho_0-\rho_{c}\right) $) to achieve
that saturation.  Figure~\ref{fig:DynamicsCylinder} reveals an excellent correspondence between the analytical
results and their numerical counterparts as explored using our
finite element implementation.  

Although it tests a subset of our full surface formulation due to the highly symmetric geometry and initial condition that make this analytic solution tractable, 
we view the correspondence between the analytical
results and the finite element solution as a helpful validation of our formulation and numerical
approach, which uses the full projection machinery of the previous sections rather than a knowledge of vector calculus in cylindrical coordinates.

\subsection{Solving the Toner-Tu Equations on a Spherical Surface}
\label{SectionSphereAnalytical}

\begin{figure}
\centering{\includegraphics[width=6.5truein]{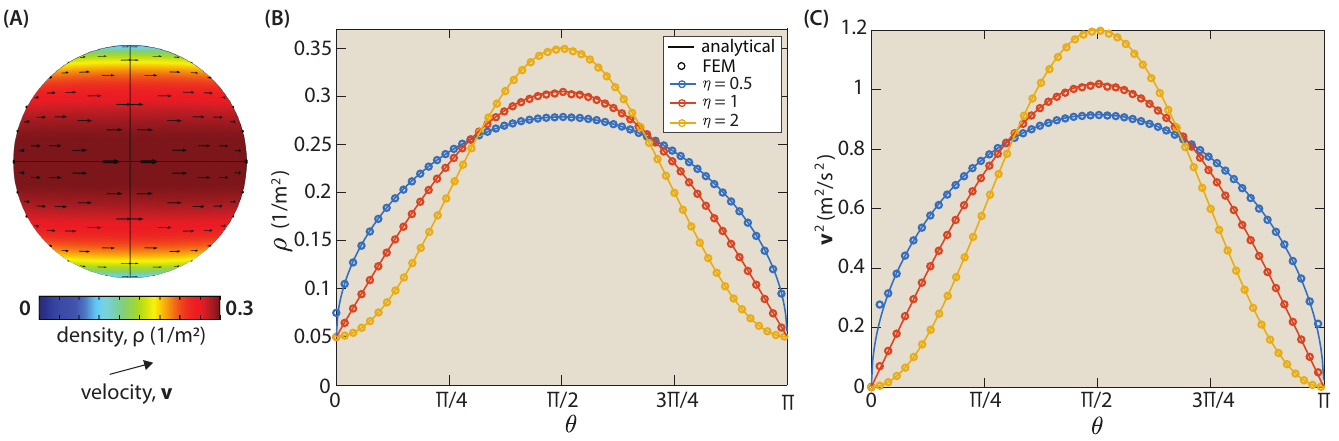}}
\caption{Comparison of analytic and numerical (COMSOL) solutions 
to the Toner-Tu equations on a sphere.
(A) Example steady-state solution for Toner-Tu herds on the surface of a sphere, for $\eta$ = 1.
(B) Steady-state herd density on the sphere. Different choices
of the dimensionless parameter $\eta$ elicit different density profiles.  (C) Steady-state velocity ${\bf v}^2$ for Toner-Tu herds on the surface of a sphere.
As in part (B), different choices of the dimensionless parameter $\eta$ result in different profiles. Parameter values used: $\alpha$ = 2 ${\mbox m^2/s}$; $\sigma$ = 40 ${\mbox m^4/s^2}$, 20 ${\mbox m^4/s^2}$, or 10 ${\mbox m^4/s^2}$, respectively; others as in Section \ref{Section:ParametersSection}. In both (B) and (C), analytical
results (curves) are in good accord with numerical results (open circles) obtained
using the finite element method. 
\label{fig:rhossCOMSOLvsAnalytical}}
\end{figure}

We next take up an analysis of the minimal Toner-Tu model on the sphere.
Previous work \cite{ Shankar2017, Sknepnek2015} described a highly-symmetric rotating band solution for flocks on the sphere, in which both the density and velocity depend only upon the polar angle $\theta$ at steady-state and not upon the azimuthal angle $\phi$.  
Here, we present the analytical solution for density and velocity on the sphere derived by Shankar {\it et al.} \cite{Shankar2017}, for comparison to our numerical results.
For their steady-state analysis, Shankar {\it et al.} \cite{Shankar2017} make the approximation that the neighbor coupling term with coefficient $D$ is absent, resulting
in a simpler minimal version of the steady-state Toner-Tu equations of the form
\begin{equation}
\lambda v^{\nu} \nabla_{\nu} v^{\mu}=\left[\alpha \left(\rho-\rho_{c}\right)-\beta g_{a b} v^{a} v^{b}\right] v^{\mu}-\sigma \nabla^{\mu} \rho.
\label{eqn:TonerTuSphereSS}
\end{equation}
We note that while Shankar {\it et al.} chose the field variables $\rho$ and ${\bf p}=\rho {\bf v}$, we continue in the language of  $\rho$ and ${\bf v}$.
In Appendix Section \ref{SectionAppendixSphereAnalytical}, following Shankar {\it et al.}, we present a step-by-step derivation of the steady-state solution of eqn. \ref{eqn:TonerTuSphereSS} using the spherical surface parameterization
${\bf r}(\theta, \phi)=(R \cos \phi \sin \theta, R \sin \phi \sin \theta, R \cos \theta)$.
This analytical calculation leads to the
 steady-state solution for flock or herd density on the sphere,
\begin{equation}
\rho_{s s}(\theta)=\rho_{c}+\left(\rho_{0}-\rho_{c}\right) A_{\eta} \sin ^{\eta} \theta,
\end{equation}
where $\eta$ is a dimensionless parameter given by
\begin{equation}
\eta=\frac{\lambda \alpha}{\beta \sigma}
\end{equation}
and the prefactor $A_{\eta}$ is defined by
\begin{equation}
A_{\eta}=2 \Gamma((3+\eta) / 2) /[\sqrt{\pi} \Gamma(1+\eta / 2)],
\end{equation}
and $\Gamma$ is the gamma function.
Note that the dimensionless parameter $\eta$ gives us a convenient knob to tune in our calculations, which can be compared directly with the numerical results of our finite element calculations
as shown in Figure~\ref{fig:rhossCOMSOLvsAnalytical}(B).  
The analytic solution and the result of our finite element treatment of the same problem agree convincingly.  We view this as a crucial validation that explores the parameter 
dependence of the solution and tests that the finite element treatment
of the various curved-space derivatives is done correctly.

In Appendix Section \ref{SectionAppendixSphereAnalytical}, following the work of Shankar {\it et al.} \cite{Shankar2017}, we also derive an analytical solution for steady-state herd velocity on the sphere as
\begin{equation}
\left|\mathbf{v}_{s s}\right|^{2}=\frac{\alpha}{\beta}\left(\rho_{0}-\rho_{c}\right) A_{\eta} \sin ^{\eta} \theta.
\end{equation}
For the specific case of $\eta=2$, for example, we have
\begin{equation}
A_2={2 \Gamma({5 \over 2}) \over \sqrt{\pi} \Gamma(2)} = {3 \over 2}.
\end{equation}
Using this value for $A_2$, we can write the density and the velocity fields as
\begin{equation}
\rho_{ss}(\theta)=\rho_c+{3 \over 2} (\rho_0-\rho_c) \mbox{sin}^2\theta
\end{equation}
and
\begin{equation}
\left|\mathbf{v}_{s s}\right|^{2}={3 \over 2} (\rho_0-\rho_c) ~\mbox{sin}^2\theta,
\end{equation}
respectively. These solutions from the work of Shankar {\it et al.} \cite{Shankar2017} provide an important opportunity to test our numerical
implementation of the Toner-Tu theory on a curved surface,
as shown in Figure~\ref{fig:rhossCOMSOLvsAnalytical}(C) for $|{\bf v}_{ss}|^2$.
As for the density, once again the correspondence between the analytic
results and the finite element treatment of the problem is convincing.  
\section{Wildebeest on Spheres and Hills: Complex Flocking Dynamics}
\label{section:BiologicalGeometries}

In the remainder of the paper, we move beyond situations where analytical solutions are feasible, using the curved-space Toner-Tu formulation and the finite element method to explore more complex herding dynamics and herding behaviors on more complex geometries.
\begin{figure}[H]
	\centering{\includegraphics[width=6.5truein]{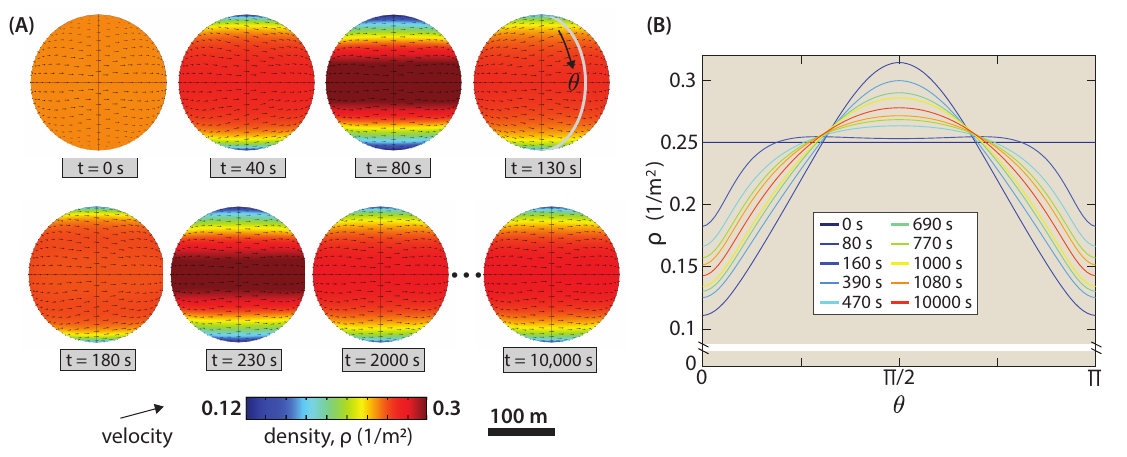}}
	\caption{Toner-Tu dynamics on the sphere.  (A) Dynamics of herd density and velocity on the sphere, during the evolution from an initial condition of uniform density  and circumferential velocity $v_{\phi} = v_{\mbox{\footnotesize{pref}}} \sin \theta$ toward the steady-state band pattern shown at 2,000 s and 10,000 s. In the long-time limit, density oscillations are damped out and a steady state emerges.  (B) Density as a function of angular position, showing the damping of density oscillations over time. Parameters as defined in Section~\ref{Section:ParametersSection}. See also Video 2.
		\label{fig:SphereDynamics}}
\end{figure}

\subsection{Toner-Tu Dynamics on the Sphere}
\begin{figure}
\centering{\includegraphics[width=5.8truein]{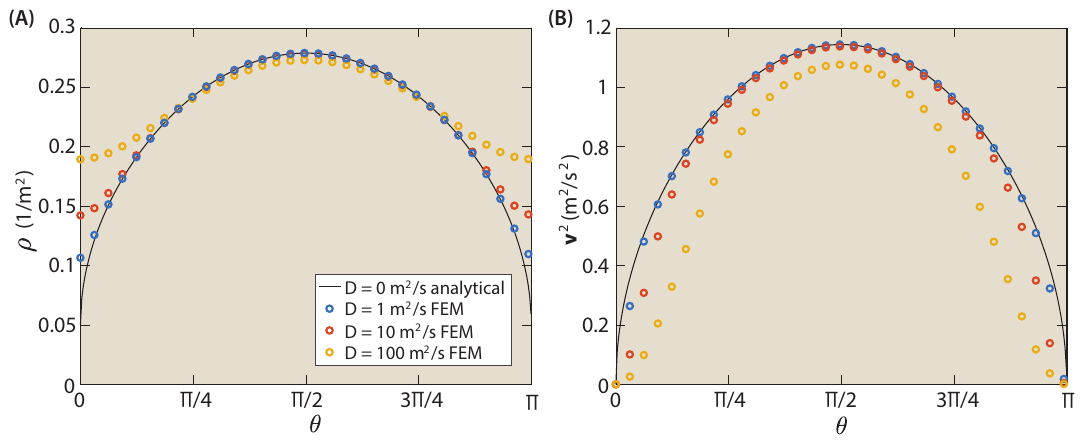}}
\caption{Steady-state solutions on the sphere including the neighbor coupling term.  (A) Steady-state
density as a function of angular position $\theta$ for different choices of the neighbor coupling
coefficient $D$.  An analytic solution in the absence of the  $D$ term is plotted for comparison.  (B) Steady-state velocity squared as a function of angular position $\theta$ for the same parameters values and analytic solution as shown in (A).
\label{fig:SphereDiffusion}}
\end{figure}

First, we continue with the case study of
herding on the sphere but take a step beyond the analytic solution provided in Section \ref{Section:AnalyticSolutions}. We add back in the analytically intractable neighbor coupling term $D\nabla^2 {\bf v}$ and examine its contribution.
In other words, we explore the dynamics of density and velocity on the spherical surface using the full finite element formulation of Section \ref{Section:FiniteElement}, including all the terms 
present in eqn.~\ref{eqn:TonerTuFEMComplete}.  
We begin with uniform density and a circumferential
velocity field $v_{\phi}= v_{\mbox{\footnotesize{pref}}} \sin \theta$.  As seen in Figure~\ref{fig:SphereDynamics} and Video 2, this solution
develops oscillations in which a nonuniform density moves back and forth in the $\theta$-direction, from the north and south poles to the equator.  The presence of the neighbor coupling term damps out 
these oscillations as shown in Figure~\ref{fig:SphereDynamics}(B). We qualitatively observe
that smaller $D$ corresponds to a longer damping time.
In the long-time limit, damping by the neighbor coupling term enables the emergence of
a steady state.  
Figure~\ref{fig:SphereDiffusion} shows several of these steady-state solutions for different choices
of the neighbor coupling coefficient $D$. We note that at small $D$, the solution approaches
the steady-state analytic solution which is attainable in the absence of this term.

\begin{figure}
\centering{\includegraphics[width=5.8truein]{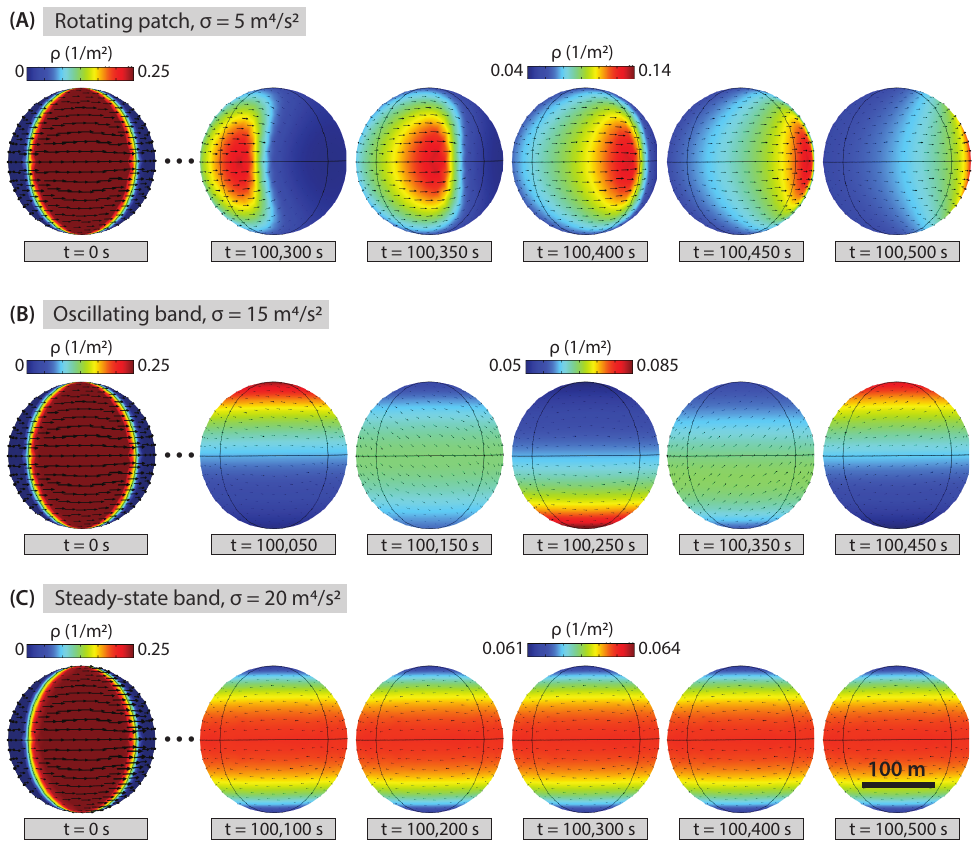}}
\caption{Herds on the sphere can form rotating patch, oscillating band, and steady-state band patterns. In these examples, herd dynamics are seeded at t = 0 s by confining wildebeest to an angular wedge between $\phi=0$ and $\phi=\pi/2$ and giving them a circumferential velocity $v_{\phi}= v_{\mbox{pref}} \sin\theta$. (A) Given a smaller pressure term ($\sigma = 5 ~{\mbox m^4/s^2}$), a rotating patch pattern of herd density emerges. (B) A larger pressure term ($ \sigma = 15 ~{\mbox m^4/s^2}$) leads the density field to exhibit temporal oscillations about the equatorial plane ($\theta = \pi/2$),
while remaining axisymmetric in $\phi$. (C) Given an ever larger pressure term  ($ \sigma = 20 ~{\mbox m^4/s^2}$), herd density is restricted to a narrow range around $\rho_{0} ~(0.0625 ~{\mbox m^{-2}})$, and a steady-state banded pattern of density emerges. See also Video 3.
\label{fig:SpherePhaseSep}}
\end{figure}

\begin{figure}
	\centering{\includegraphics[width=4.6truein]{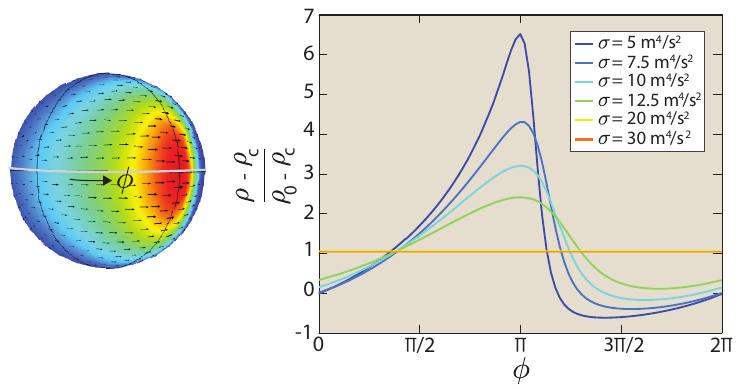}}
	\caption{Increasing the Toner-Tu pressure term drives herd density from patch to band patterns. Plot shows density as a function of $\phi$, angular position around the sphere circumference. The simulation was seeded with the initial condition depicted in Figure \ref{fig:SpherePhaseSep} for varied $\sigma$ that lead to rotating patch or steady-state band solutions in the long-time limit. Curves are aligned such that their maxima are at $\phi = \pi$. 
		\label{fig:SphereRhoPhiPlot}}
\end{figure}

However, the dynamics that can take place on the spherical surface can be significantly richer than the steady-state band solution shown
above.  To seed the dynamics, we consider an initial condition in which the active agents are confined
to an angular wedge between $\phi=0$ and $\phi=\pi/2$, again with circumferential velocity $v_{\phi}= v_{\mbox{\footnotesize{pref}}} \sin~\theta$, as shown in
Figure~\ref{fig:SpherePhaseSep} and Video 3. 
This asymmetric initial condition can give rise to both dynamic and steady-state solutions depending on the value of parameters such as $\rho_{0}$ or $\sigma$. In Figure~\ref{fig:SpherePhaseSep}, different choices of $\sigma$ lead to different herding behaviors in the long-time limit. In Figure~\ref{fig:SpherePhaseSep}(A), at small $\sigma$, a droplet-shaped patch of dense wildebeest circulates around the sphere indefinitely. Tuning $\sigma$ to increase the magnitude of the pressure term leads instead to the emergence of oscillating density bands, as shown in Figure~\ref{fig:SpherePhaseSep}(B), and then to steady-state density bands, as shown in Figure~\ref{fig:SpherePhaseSep}(C). Note that we have not highlighted the interesting 
transient dynamics displayed as herds evolve from the wedge initial condition toward the patch and band patterns. For the rotating patch and steady-state band solutions, which reach dynamic or steady-state solutions with mirror symmetry across the equatorial plane ($\theta =\pi/2$), we can plot herd density as a function of $\phi$ to characterize the shape of the density patch and to further explore the effect of increasing the Toner-Tu pressure term. As shown in Figure~\ref{fig:SphereRhoPhiPlot}, increasing $\sigma$ drives the emergent density pattern from a patch to a band by flattening $\rho(\phi)$ from the interesting asymmetric patch shape observed at $\sigma$ = 5 ${\mbox m^4/s^2}$ to the uniform profile observed for $\sigma$ = 20 ${\mbox m^4/s^2}$ and $\sigma$ = 30 ${\mbox m^4/s^2}$.

\subsection{Herding over a Gaussian Hill}
\begin{figure}[h]
	\centering{\includegraphics[width=6.5truein]{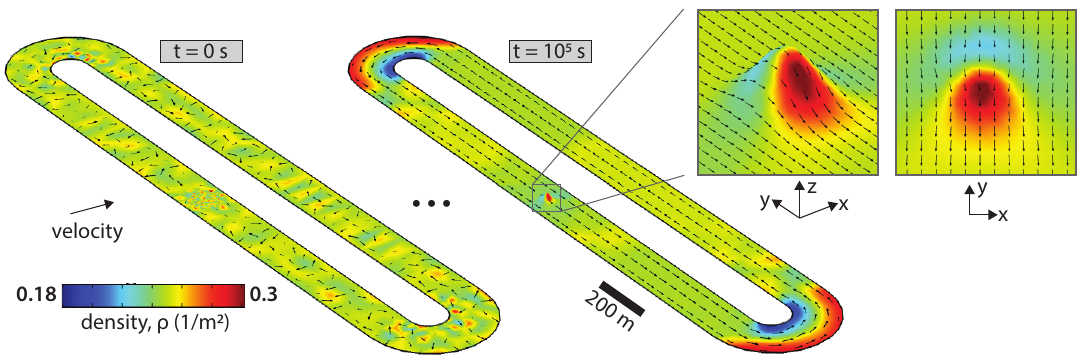}}
	\caption{Curvature induces local density and velocity changes. Steady-state herd density (colors) and velocity (black arrows) for wildebeest traversing a Gaussian hillside embedded in a racetrack straightaway. From an initial condition of random velocity orientation, a steady-state recirculation around the racetrack emerges in the long-time limit. In an effect reminiscent of lensing \cite{Kamien2009, Green2017}, the Gaussian hill's curvature induces low and high density regions.
		\label{fig:RacetrackHill}}
\end{figure}

Moving beyond the symmetric geometry of the sphere, we next explore the contribution of curvature to steady-state density and velocity for a herd on a racetrack. As shown in Figure \ref{fig:RacetrackHill}, the addition of a Gaussian hill on the racetrack straightaway leads to local changes in herd density. Density is lower on the face of the hill approached by the wildebeest, as they climb and their velocity field diverges, increasing the space between wildebeest neighbors as their paths bend around the hill. Density is higher on the face of the hill that the wildebeest descend, where the velocity field again converges. This curvature-dependent effect is loosely reminiscent of lensing \cite{Kamien2009, Green2017}, but we have not rigorously explored the analogy. 

\subsection{Herding on an Undulating Island Landscape}
In a final case study, inspired by wildebeest herding on complicated natural landscapes, we consider the motion of a Toner-Tu herd on an island of Gaussian hillsides. We offer this study as a step toward a dialogue with real-world data, for which we hope the approach presented in this paper may prove useful. For this final exploration of how herding behavior is altered by the local topography of a landscape, we develop one additional term for the herding equations that govern the time evolution of wildebeest velocity.

\begin{figure}
\centering{\includegraphics[width=4.2truein]{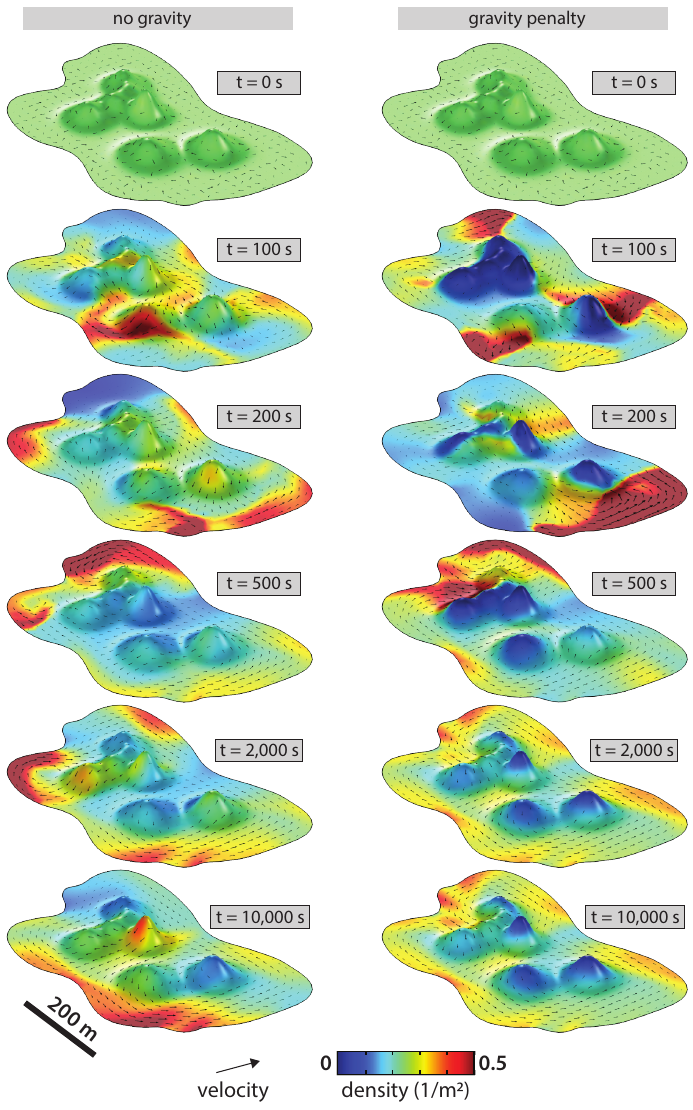}}
\caption{Dynamics of a Toner-Tu herd on an undulating landscape, initialized with uniform density $\rho_{0} = 0.25~ {\mbox m^{-2}}$ and randomly oriented velocities. Comparing the left column ($\zeta = 0~ {\mbox m/s^2}$) and the right column ($\zeta = 0.05~ {\mbox m/s^2}$) highlights the contribution of the gravitational force. In the presence of the gravitational force, the hills act as soft obstacles, favoring herd circulation around them. Other parameters used were presented in Section ~\ref{Section:ParametersSection}. See also Video 4.
\label{fig:WildebeestIsland}}
\end{figure}

In much the same way that body forces are incorporated into the
Navier-Stokes equations, we hypothesize that the topography of
the landscape upon which the wildebeest are marauding acts upon them through a gravitational force term of the form
\begin{equation}
{\bf f}_g= -\zeta {\bf k},
\end{equation}
where {\bf k} is the unit vector in the z-direction (height). We note that the units of $\zeta$ are not strictly those of force
since the  Toner-Tu equations simply have dimensions of acceleration.
In the same spirit that we selected the other parameters in Section \ref{Section:ParametersSection} to ensure terms of comparable
in magnitude, we estimate that $\zeta$ is on the order of $0.1~\mbox{m}/\mbox{s}^2$. For the simulations shown in Figure~\ref{fig:WildebeestIsland}, we chose $\zeta = 0.05~\mbox{m}/\mbox{s}^2$
to prevent the emergence of unphysical negative densities at hilltops. To write the curved-surface contribution of the gravitational force term to the dynamical equations,
we carry out the surface projection as usual, 
\begin{equation}
{\bf P}{\bf f}_g= -\zeta {\bf P}{\bf k} = -\zeta {\bf k}+\zeta  {\bf n} ({\bf n} \cdot {\bf k}),
\end{equation}
where we recognize that $({\bf n} \cdot {\bf k})=n_3$.  
Because of the projection operator, this term affects all three components of
velocity. For example, we can write the Toner-Tu equation governing the 1-component of velocity, with the addition of the gravitational force, as
\begin{equation}
{\partial  v_1^{\parallel} \over \partial t}=[\alpha (\rho - \rho_c) - \beta v_j^{\parallel} v_j^{\parallel}]v_1^{\parallel}
-\sigma \left(\frac{\partial \rho}{\partial x_{1}}\right)_{\Gamma} +D P_{1l} \left(\frac{\partial G_{l j}}{\partial x_{j}}\right)_{\Gamma} -\lambda v_j^{\parallel} \left[\left({\partial  v_1^{\parallel}\over \partial x_j}\right)_{\Gamma}-n_1n_k \left({\partial  v_k^{\parallel}\over \partial x_j}\right)_{\Gamma} \right]+\zeta n_1 n_3.
\end{equation}
As in the channel flow example, where the preferred speed term and the neighbor coupling term compete to modify velocity magnitude near no-slip boundaries, the preferred speed and the gravitational term compete on sloped hillsides.

To explore the dynamics of an active herd on this undulating
landscape, we seed the region with an initial distribution of
wildebeest at uniform density $\rho_0$ and with random velocity orientation.
From this initial distribution, over time the wildebeest self-organize into a herd that navigates around the complex landscape. In Figure~\ref{fig:WildebeestIsland} and Video 4, we show an example of these dynamics and explore the contribution of the gravitational force term. In the presence of the gravitational force, the hills effectively serve as ``soft" obstacles, directing the herd to circulate around them. These obstacles can stabilized fixed patterns in the velocity field, as on the right in Figure \ref{fig:WildebeestIsland} and Video 4.

We hope that these playful Toner-Tu solutions on the complex geometry of a hill-studded island show the versatility of the approach presented in this work. We also hope that they visually illustrate a vision of the dialogue between theory and real-world data towards which we ultimately strive, and towards which the current study takes one step.

\section{Conclusion}

The collective motion of animals, cells, and molecules presents a beautiful phenomenon. Toner-Tu flocking theory has proven a versatile mathematical description of such phenomena, helping to enable an expansion from traditional continuum models of materials into the relatively uncharted waters of the dynamics of living matter. 
Toner-Tu and other continuum active matter models make possible a dialogue between mathematical prediction and the stunning living world outside our windows.
To undertake such a dialogue necessitates solving these equations for real-world geometries and boundary conditions. Flocks of sheep, collectively migrating cells, collectively flowing cytoskeletal filaments -- all of these flocking agents move within structured environments that must be accounted for to make a comparison between \textit{in vivo} data and predictions of theoretical models. 

Here, we focused on flocking phenomena that occur on curved surfaces, such as wildebeest navigating terrain, or the motility-driving flows of cytoskeletal filaments at the surface of protozoan parasites of particular interest to us \cite{Hueschen2022c}. Both the formulation and solution of the Toner-Tu equations is more challenging once our goal is their implementation on curved surfaces. Frequently, surfaces of biological interest are highly asymmetric, rendering analytical approaches challenging or impossible. In this work we sought to present, in accessible form, a general curved-surface formulation of the Toner-Tu equations and its finite element method implementation. We hope that our pedagogical derivation of a surface formulation using the tools of extrinsic differential geometry, and our explication of numerically solving it on arbitrary curved surfaces, will prove useful to others interested in a versatile approach for exploring continuum theories on complex shapes. We also sought to demonstrate the equivalence of our approach to analytical ones, for simple geometries where analytical results are licensed. Finally, we sought to harness the versatile power of this approach by exploring its predictions across a range of geometries, boundary conditions, and parameter values. By observing the effect of geometry and parameter choice on herd direction reversal in a channel with a circular obstacle in Figure \ref{fig:ChannelFlowObstacle2}, for example, we demonstrated the ability of this approach to generate hypotheses that can subsequently be explored analytically, as in  Figure \ref{fig:ChannelFlowObstacle3}. By exploring the local density changes induced by curvature for herds passing over a Gaussian hill in Figure \ref{fig:RacetrackHill}, we highlighted the ability of this approach to generate intriguing observations that generate new questions. By predicting the dynamics of herds on a undulating island landscape in Figure \ref{fig:WildebeestIsland}, we took a step towards a vision of predictive dialogue with data from the rich, structured environment in which we live.

Looking forward, we dream of increased dialogue between continuum active matter theory and measurements taken in our rich living world, from drone footage of migrating animal herds to movies of microscopic cytoskeletal flows within embryos and cells. Such a dialogue will enable the determination of absolute parameter values for such systems, which will in turn make possible the contrivance of experiments designed to test the theory. 
Can the same field theories, indeed, be applied to sheep and to microscopic microtubules? What is the effect of curvature on velocity for different types of agents? We hope that our presentation of the approach taken here will contribute towards efforts to bridge active matter theory and complex real-world shapes, directly testing whether our observations of the living world can be understood and predicted in the language of continuum theories of active collectives.

\section{Appendix}

\subsection{Calculus on Surfaces Via Projection Operators}

\label{Section:VectorTensor}

The formulation of the coupled partial differential equations that describe the herd density
and velocity fields required us to perform various derivative operations on
surfaces.  To do so, we used extrinsic differential geometry to take the tangential calculus approach
described pedagogically by Jankuhn {\it et al.}~\cite{Jankuhn2018} and in 
Tristan Needham's recent beautiful book~\cite{Needham2020}.  This approach involves defining all necessary differential operators in terms of the projection operator, ${\bf P} = {\bf I}-{\bf n} \otimes {\bf n}$ or $P_{ij} = \delta_{i j}-n_{i} n_{j} $, instead of working in the language of parameterized
surfaces. In this appendix, we set these tangential calculus operations front and center. We define the tangential curved-space differential operators used in this work, and we attempt to convey some intuition
for what these operations achieve.\\

\noindent {\it Gradient of a Scalar}\\
Our treatment of Toner-Tu surface problems requires us to compute derivatives
in the tangent plane of the surface of interest. For example, we might be interested in the gradient of 
a scalar field such as the density $\rho({\bf r})$. 
We can think of the surface gradient as resulting from projecting the full 3D gradient onto the tangent
plane, using the projection operator ${\bf P}$  defined in eqn.~\ref{eqn:ProjectionOperator}. Thus, we find
\begin{equation}
\begin{aligned}
{\bf P} \nabla \rho = ({\bf I}-{\bf n} \otimes {\bf n}) \nabla \rho &=\left(\delta_{i j}-n_{i} n_{j}\right) \frac{\partial \rho}{\partial x_{j}} \\
&=\frac{\partial \rho}{\partial x_{i}}-n_{i} n_{j} \frac{\partial \rho}{\partial x_{j}}.
\end{aligned}
\end{equation}
Hence, throughout the paper, whenever we need to invoke the tangent curved-space gradient operator, we do so
in the form
\begin{equation}
\left(\nabla_{\Gamma} \rho\right)_{i}=(\nabla \rho)_{i}-n_{i} n_{j} \frac{\partial \rho}{\partial x_{j}} =\frac{\partial \rho}{\partial x_{i}}-n_{i} n_{j} \frac{\partial \rho}{\partial x_{j}}.
\end{equation}\\

\noindent {\it Gradient of a Vector}\\
Several terms in the minimal Toner-Tu model involve more complex differential operations.
For example, the advection term in the Toner-Tu theory
requires us to evaluate the gradients of the velocity field (a vector!).   Thus, we also need to formulate the mathematical tools for constructing the gradient
of a vector. In direct notation, the tangential curved-space gradient of a vector is
\begin{equation}
\nabla_{\Gamma} \mathbf{v} =\mathbf{P}(\nabla \mathbf{v}) \mathbf{P} ,
\end{equation}
which we can write in indicial notation as
\begin{equation}
\left(\nabla_{\Gamma} {\bf v}\right)_{ij} 
= [ {\bf P} (\nabla {\bf v}) {\bf P} ]_{ij}
=P_{ik} {\partial v_k \over \partial x_l}  P_{lj}.
\end{equation}
We can expand this out using the definition of the projection operator resulting in
\begin{equation}
P_{ik} {\partial v_k \over \partial x_l}  P_{lj}=(\delta_{ik} - n_in_k) {\partial v_k \over \partial x_l}
(\delta_{lj} - n_ln_j) 
\end{equation}
which simplifies to the form
\begin{equation}
P_{ik} {\partial v_k \over \partial x_l}  P_{lj}=\left(\frac{\partial v_{i}}{\partial x_{j}}-n_ln_{j} \frac{\partial v_{i}}{\partial x_{l}}\right) - n_in_k \left(\frac{\partial v_{k}}{\partial x_{j}}-n_ln_{j} \frac{\partial v_{k}}{\partial x_{l}} \right),
\end{equation}
an expression that we will see repeatedly.
We note that the final result features the projected gradient of the $v_i$ and $v_k$ components of
velocity, respectively.\\

\noindent {\it Divergence of a Vector}\\
Using the definition of the gradient of a vector presented above, we can formally write the tangential curved-space
\textit{divergence} of a vector as
\begin{equation}
\operatorname{div}_{\Gamma} \mathbf{v}=\operatorname{tr}\left(\nabla_{\Gamma} \mathbf{v}\right)=\operatorname{tr}(\mathbf{P}(\nabla \mathbf{v}) \mathbf{P})=\operatorname{tr}(\mathbf{P}(\nabla \mathbf{v}))=\operatorname{tr}((\nabla \mathbf{v}) \mathbf{P}).
\end{equation}
The final two expressions follow from the fact that $tr({\bf ABC})=tr({\bf BCA})$ and all other cyclic permutations
as well, in conjunction with the fact that ${\bf P}^2={\bf P}$.
In component form, this simplifies to
\begin{equation}
\operatorname{div}_{\Gamma} v_{i}=\frac{\partial v_{i}}{\partial x_{i}}-n_{i} n_k\frac{\partial v_{i}}{\partial x_{k}},
\label{eqn:DivergenceVectorIndicial}
\end{equation}
an expression we used repeatedly in our derivations of the field equations and in their formulation numerically within COMSOL.\\

\noindent {\it Divergence of a Tensor}\\
As seen in the main text, the Toner-Tu theory requires us to evaluate the divergence of tensorial quantities
on curved surfaces.  For example, we invoked the stress-like quantity ${\bf J}$,
which in the full three-dimensional setting we write as
\begin{equation}
\nabla \cdot {\bf J}=J_{ji,j}.
\end{equation}
This result needs to be amended in the context of an arbitrary surface. In order to define the divergence of a tensor, we can examine separately the divergence of
the component vectors of that tensor, which we obtain by having the tensor act on the unit vectors. 
More precisely, we work out the divergence of a tensor ${\bf A}$, following Jankuhn {\it et al.} \cite{Jankuhn2018},
as
\begin{equation}
\operatorname{div}_{\Gamma} \mathbf{A}=\left(\operatorname{div}_{\Gamma}\left(\mathbf{e}_{1}^{T} \mathbf{A}\right),  \operatorname{div}_{\Gamma}\left(\mathbf{e}_{2}^{T} \mathbf{A}\right), \operatorname{div}_{\Gamma}\left(\mathbf{e}_{3}^{T} \mathbf{A}\right)\right)^{T}
\label{eqn:DivergenceTensorDirect}
\end{equation}
where we use the definitions
\begin{equation}
{\bf e}_1^T =(1, 0, 0)
\end{equation}
\begin{equation}
{\bf e}_2^T =(0, 1, 0)
\end{equation}
and
\begin{equation}
{\bf e}_3^T =(0, 0, 1).
\end{equation}
We recall that the divergence of a tensor gives rise to a vector, consistent with the multiple components in our expression for $\operatorname{div}_{\Gamma} \mathbf{A}$.
Indeed, we can rewrite this expression in component form as
\begin{equation}
(\operatorname{div}_{\Gamma} \mathbf{A})_i = P_{lk}{\partial A_{il}\over \partial x_k}={\partial A_{ik} \over \partial x_k} - n_ln_k {\partial A_{il}\over \partial x_k}.
\end{equation}\\

\noindent  {\it Laplacian of a Vector}	\\
As shown by Jankuhn {\it et al.}~\cite{Jankuhn2018}, the surface-projected version of the Laplacian term (see their  eqn.~3.16 for the surface Navier-Stokes equations) is given by
\begin{equation}
\underbrace{\nabla^2{\bf v}}_{\mbox{in the plane}} = \underbrace{\operatorname{div}( \nabla {\bf v})}_{\mbox{in the plane}} ~\longrightarrow ~~~\underbrace{{\bf P} \operatorname{div}_{\Gamma}(\nabla_{\Gamma} {\bf v})}_{\mbox{on a curved surface}}.
\end{equation}
We will now unpack this expression for the curved surface Laplacian, from right to left.
We must first compute the
surface gradient of the vector {\bf v}, given by the tensor  ${\bf G}= \nabla_{\Gamma} {\bf v}$.
In component form, we write this as
\begin{equation}
\nabla_{\Gamma} {\bf v} = G_{ij}= \left(\frac{\partial v_{i}}{\partial x_{j}}-n_ln_{j} \frac{\partial v_{i}}{\partial x_{l}}\right) - n_in_k \left(\frac{\partial v_{k}}{\partial x_{j}}-n_ln_{j} \frac{\partial v_{k}}{\partial x_{l}} \right).
\end{equation}
We then
invoke the definition of the surface-projected divergence of the tensor ${\bf G}$,
\begin{equation}
\operatorname{div}_{\Gamma}\left({\bf G}\right)_i=P_{lk}{\partial G_{il}\over \partial x_k}.
\end{equation}
Using the definition of the projection operator, we can expand this out as
\begin{equation}
P_{lk}{\partial G_{il}\over \partial x_k}=(\delta_{lk}-n_ln_k) {\partial G_{il}\over \partial x_k}={\partial G_{il}
\over \partial x_l} -n_ln_k {\partial G_{il}\over \partial x_k}.
\end{equation}
To complete our calculation of the curved-space Laplacian of a vector, we need to compute
${\bf P} \operatorname{div}_{\Gamma}\left({\bf G}\right)$, which we can write in indicial notation
as 
\begin{equation}
P_{il} \operatorname{div}_{\Gamma}\left({\bf G}\right)_i = P_{il} \left( 
{\partial G_{lj}
\over \partial x_j} -n_jn_k {\partial G_{lj}\over \partial x_k}\right).
\label{eqn:PDivergenceG}
\end{equation}

\subsection{Implementation in COMSOL Multiphysics~\textregistered}
\label{SectionAppendixCOMSOL}
Our COMSOL Multiphysics\textregistered\, files are available for download, use, and adaptation at \url{https://github.com/RPGroup-PBoC/wildebeest_herds}. They are accompanied by a ``How-To" document that provides a tutorial on navigating and using these files.

We hope that others will find our approach user-friendly and adaptable for solving the Toner-Tu equations on other complex curved surfaces, for solving other continuum equations on curved surfaces, or for additional study of the cases presented here. For our finite element method (FEM) calculations, we chose to use the off-the-shelf software COMSOL Multiphysics\textregistered\, for its accessibility and learner-friendly interface. We note with regret, however, that the use of this software and our code requires access to a paid COMSOL Multiphysics\textregistered\, license. If you do not already have a license and are affiliated with an institution, we recommend looking into access options through a shared software library.

Below, for the convenience of COMSOL users among our readers,
we highlight some of the COMSOL-specific features and nomenclature that we used.

To solve our custom surface partial differential equations, we used the COMSOL Multiphysics\textregistered ~General Form Boundary PDE interface and took advantage of COMSOL's built-in tangential differentiation operator, dtang(f,x), discussed in detail below. 
We also made use of the normal vector (nx, ny, nz), a built-in geometric variable. 
For a thorough and practical introduction to the finite element method, we recommend reference \cite{COMSOLMultiphysics2017}. 
For simulations initialized with a disordered velocity field, every node's velocity orientation was drawn randomly from a uniform distribution of angles between 0 and 2$\pi$ and had magnitude $v(0)$. Unless noted otherwise, as in the case of the cylinder dynamics shown in Figure \ref{fig:DynamicsCylinder}, $v(0)$ = $v_{\mbox{pref}}$= 1 m/s. Other parameters are defined and discussed in detail in Section \ref{Section:ParametersSection}. On curved surfaces, to avoid the accumulation of out-of-plane components in ${\bf v}$ from numerical error, we implemented a weak constraint of ${\bf n \cdot v} = 0$. Similarly, we implemented a global constraint on the total integrated density $\rho$ on the surface, ensuring it stayed at its initial value. Default COMSOL solvers and settings were used: implicit backward differentiation formula (BDF) for time stepping and multifrontal massively parallel sparse direct solver (MUMPS) for the linear direct spatial solver. Our meshes employed between 1310 and 6525 triangular elements (1310-1674 for the channel with obstacle; 1312 for the cylinder; 6525 for the sphere; 1604 for the racetrack with hill; 2642 for the island) and can be viewed in the source files available at \url{https://github.com/RPGroup-PBoC/wildebeest_herds}.

We now provide additional details on formulating our custom equations in order to solve them in COMSOL.
The user-defined partial differential equation is written in COMSOL as
 \begin{equation}
 \frac{\partial \mathbf{v}}{\partial t}+\nabla 
\cdot {\bf \Gamma}^{(v)}={\bf f^{(v)} }.
\end{equation}
 Here,  the whole formulation comes
down to the definitions of ${\bf \Gamma}^{(v)}$ and ${\bf f^{(v)} }$. 
The flux-like quantity ${\bf \Gamma}^{(v)}$ is a $3 \times 3$ matrix that is defined such
that 
\begin{equation}
(\nabla 
\cdot {\bf \Gamma}^{(v)})_i = {\partial \Gamma_{ji}^{(v)} \over \partial x_j}.
\end{equation}
In this section, we use the notation ${\bf \Gamma}$ and ${\bf f} $ to precisely match the notation of COMSOL's General Form PDE interface. We note that in the main body of the paper we referred to ${\bf J}^{(v)}$ rather
than ${\bf \Gamma}^{(v)}$ in order to avoid confusion with subscript $\Gamma$, which signifies ``surface," following the notation of  Jankuhn {\it et al.}~\cite{Jankuhn2018} in
their description of the tangential calculus. There, we wanted to give readers the opportunity
to make contact with Jankuhn {\it et al.} with minimal difficulty.

COMSOL is constructed to allow us to evaluate what we have earlier called $\nabla_{\Gamma}$ 
using an operation known as ``dtang(f,$x_i$),'' which is the tangent plane component of the gradient
in the $i^{th}$ direction of the function $f(x_1,x_2,x_3)$.  
We can write this formally as
\begin{equation}
\nabla_{\Gamma} f=\nabla f-{\bf n} ({\bf n} \cdot \nabla f) = (dtang(f,x_1), dtang(f,x_2), dtang(f,x_3))
\end{equation}
Thus, these terms  can be written in component form as
\begin{equation}
dtang(f,x_1)= (\nabla_{\Gamma}f)_1=\frac{\partial f}{\partial x_{1}}-n_{1} n_{j} \frac{\partial f}{\partial x_{j}}.
\end{equation}
Specifically, we have for the
gradient in the ``1-direction''
\begin{equation}
dtang(f,x_1)= { \partial f \over \partial x_1}- n_1(n_1{ \partial f \over \partial x_1}+n_2{ \partial f \over \partial x_2}+n_3{ \partial f \over \partial x_3}),
\end{equation}
with similar results for the other directions given by
\begin{equation}
dtang(f,x_2)= { \partial f \over \partial x_2}- n_2(n_1{ \partial f \over \partial x_1}+n_2{ \partial f \over \partial x_2}+n_3{ \partial f \over \partial x_3})
\end{equation}
and
\begin{equation}
dtang(f,x_3)= { \partial f \over \partial x_3}- n_3(n_1{ \partial f \over \partial x_1}+n_2{ \partial f \over \partial x_2}+n_3{ \partial f \over \partial x_3}).
\end{equation}

To provide detailed examples for those interested in similar FEM implementations, we now translate key expressions from our curved-space Toner-Tu formulation into COMSOL syntax. In eqn.~\ref{eqn:GMatrix} we defined the  matrix $G_{ij}$
which can be written in COMSOL syntax as
\begin{equation}
G_{ij} = dtang(v_i^{\parallel},x_j) - n_i n_k dtang(v_k^{\parallel}, x_j), 
\end{equation}
where $v_i^{\parallel}$ refers to the tangent plane component of the $i^{th}$ component of velocity.  For example, in expanded form,
\begin{equation}
G_{21} = dtang(v_2^{\parallel}, x_1) - n_2 \left( n_1 dtang(v_1^{\parallel}, x_1) + n_2 dtang(v_2^{\parallel}, x_1) + n_3 dtang(v_3^{\parallel}, x_1) \right). 
\end{equation}
In addition, we introduced a force term in Section \ref{Section:FiniteElement} to recapitulate the Toner-Tu equations.
When written in COMSOL format, those terms are of the form
\begin{equation}
\begin{aligned}
f_1^{(v)} + f_1^{\mbox{\mbox{fict}}} & =
 (\alpha (\rho-\rho_c)- \beta ((v_1^{\parallel})^2 + (v_2^{\parallel})^2+(v_3^{\parallel})^2))v_1^{\parallel}
- \sigma ~dtang(\rho, x)\\
&-\lambda \Big(v_1^{\parallel} ~dtang(v_1^{\parallel},x_1)+v_2^{\parallel} ~dtang(v_1^{\parallel},x_2)+
v_3^{\parallel} ~dtang(v_1^{\parallel},x_3)\\
&-v_1^{\parallel}n_1n_1 ~dtang(v_1^{\parallel},x_1)
-v_1^{\parallel}n_1n_2 ~dtang(v_2^{\parallel},x_1)
-v_1^{\parallel}n_1n_3 ~dtang(v_3^{\parallel},x_1)\\
&-v_2^{\parallel}n_1n_1 ~dtang(v_1^{\parallel},x_2)
-v_2^{\parallel}n_1n_2 ~dtang(v_2^{\parallel},x_2)
-v_2^{\parallel}n_1n_3 ~dtang(v_3^{\parallel},x_2)\\
&-v_3^{\parallel}n_1n_1 ~dtang(v_1^{\parallel},x_3)
-v_3^{\parallel}n_1n_2 ~dtang(v_2^{\parallel},x_3)
-v_3^{\parallel}n_1n_3 ~dtang(v_3^{\parallel},x_3)\Big)\\
&-D \left(dtang(P_{11},x_1)G_{11}+ dtang(P_{11},x_2)G_{12}+dtang(P_{11},x_3)G_{13}\right) \\
&-D\left(dtang(P_{12},x_1)G_{21}+ dtang(P_{12},x_2)G_{22}+dtang(P_{12},x_3)G_{23}\right)\\
&-D \left(dtang(P_{13},x_1)G_{31}+ dtang(P_{13},x_2)G_{32}+dtang(P_{13},x_3)G_{33}\right),
\end{aligned}
\end{equation}
for the $x_1-$component.  Note that  the $\alpha$ and $\beta$ terms capture the preferred speed
part of Toner-Tu theory, the term involving
$\sigma$ treats the pressure term,  the  terms preceded by $\lambda$ capture the advection contribution 
to the Toner-Tu equations, and the final pieces involving ${\bf P}$ and ${\bf G}$ subtract
off the fictitious force due to the curved-space Laplacian.
We have similar equations for the $x_2$- and $x_3$-components of the force.

We add here an additional note concerning our implementation of user-defined partial differential
equations within COMSOL using $\nabla \cdot {\bf \Gamma}$, where
the tensor ${\bf \Gamma}$ is a generalized stress-like term.  As we saw in the
context of the curved-space version of the Laplacian above, we wished to implement an equation of form
eqn.~\ref{eqn:PDivergenceG}
which is {\it not} in the form of the divergence of a tensor.  Thus,
we introduced 
\begin{equation}
\left( {\partial (P_{il}G_{lj}) \over \partial x_j}\right)_{\Gamma}= {\partial (P_{il}G_{lj}) \over \partial x_j}
- n_jn_k {\partial (P_{il}G_{lj}) \over \partial x_k}.
\end{equation}
The right side of this equation asks us to do ordinary calculus according to the product rule, culminating in
\begin{equation}
\left( {\partial (P_{il}G_{lj}) \over \partial x_j}\right)_{\Gamma}= P_{il}{\partial G_{lj} \over \partial x_j} +
{\partial P_{il} \over \partial x_j} G_{lj}- n_jn_k {\partial P_{il} \over \partial x_k} G_{lj}-n_jn_k P_{il} {\partial G_{lj} \over \partial x_k}.
\end{equation}
We can rewrite this as
\begin{equation}
\left( {\partial (P_{il}G_{lj}) \over \partial x_j}\right)_{\Gamma}= P_{il}\left({\partial G_{lj} \over \partial x_j} --n_jn_k{\partial G_{lj} \over \partial x_k}\right)+ 
\left({\partial P_{il} \over \partial x_j} - n_jn_k {\partial P_{il} \over \partial x_k}\right) G_{lj}.
\end{equation}
This proves the assertion used in the paper that  the quantity we want  can be written as the divergence of a tensor  minus an unwanted term that we can treat as a force, resulting in
\begin{equation}
P_{il} \left( {\partial G_{lj} \over \partial x_j}\right)_{\Gamma}= \left( {\partial (P_{il}G_{lj}) \over \partial x_j}\right)_{\Gamma} 
- \left({\partial P_{il} \over \partial x_j}\right)_{\Gamma} G_{lj}.
\end{equation}

\subsection{Analytical Solutions for Toner-Tu on the Cylinder}
\label{SectionAppendixCylinderAnalytical}
In this appendix section, we derive analytical solutions on the cylinder for steady-state Toner-Tu density and velocity, and for velocity dynamics during relaxation to that steady state. These solutions are used for comparison to our finite element method results in Section \ref{SectionCylinderAnalytical}.
To develop analytic intuition on a cylindrical surface in order to complement our finite 
element studies, our first step is the parameterization of the cylindrical surface,
\begin{equation}
{\bf r}(\theta, z)=(R \cos \theta, R \sin \theta, z).
\end{equation}
We have two parameters, $u^1=\theta$ and $u^2=z$, which provide an ``address'' for every point on
the cylindrical surface.  That is, every choice of $(u^1,u^2)$ has a corresponding point on the surface.
Next we need to find the tangent vectors, which are defined
via
\begin{equation}
{\bf a}_i = \frac{\partial {\bf r}}{\partial u^i},
\end{equation}
where $u^i$ is the ``coordinate'' of the surface defined above.
Given these definitions, we have 
\begin{equation}
{\bf a}_{\theta}=\frac{\partial {\bf r}}{\partial \theta} =(-R \sin \theta, R \cos \theta, 0) = R ~{\bf e}_{\theta}
\end{equation}
and
\begin{equation}
{\bf a}_z=\frac{\partial {\bf r}}{\partial z}=(0,0,1)={\bf e}_{z}
\end{equation}
as our two tangent vectors, exactly as expected.   Note that these vectors do {\it not} have the same units.  
Indeed, this issue of units is revealed in the metric tensor.
The components of the  metric tensor are obtained by computing the dot product of these
tangent vectors and are given as
\begin{equation}
g_{\alpha \beta} \equiv {\bf a}_{\alpha} \cdot {\bf a}_{\beta}.
\end{equation}
For the specific case of a cylinder, the metric tensor takes the particularly simple form
\begin{equation}
{\bf g}= \left[\begin{array}{ll}
R^{2} & 0 \\
0 & 1
\end{array}\right]
\end{equation}
with much of the simplicity resulting from the fact that the tangent vectors are orthogonal.

With this background, we now want to solve the Toner-Tu equations 
\begin{equation}
{\partial v^{\mu} \over \partial t}=\left[\alpha \left(\rho-\rho_{c}\right)-\beta g_{a b} v^{a} v^{b}\right] v^{\mu}-\sigma \nabla^{\mu} \rho-\lambda v^{\nu} \nabla_{\nu} v^{\mu},
\end{equation}
for the case of the cylinder. Following the work of Shankar {\it et al.}, we neglect the diffusive (neighbor coupling) term, both to make an analytical solution tractable and to compare directly with their previous study. 
This formulation invokes the covariant derivative
\begin{equation}
\nabla_{\mu} v^{\nu}=\partial_{\mu} v^{\nu}+\Gamma_{\alpha \mu}^{\nu} v^{\alpha}
\end{equation}
where $\Gamma_{\alpha \mu}^{\nu}$ are the  Christoffel symbols.  All of this extra machinery
and corresponding notation comes down to the subtlety of getting our derivatives
right in the curved-space setting.  We also introduce the symbol $\Delta$ to represent the intrinsic curved-space version of the Laplacian.
We assume that the Toner-Tu steady-state,
\begin{equation}
\lambda v^{\nu} \nabla_{\nu} v^{\mu} = \left[\alpha \left(\rho-\rho_{c}\right) -\beta g_{a b} v^{a} v^{b}\right] v^{\mu}-\sigma \nabla^{\mu} \rho,
\label{eqn:TonerTuSphereSSAppendixCylinder}
\end{equation}
has no velocity in the z direction.  
Eqn.~\ref{eqn:TonerTuSphereSSAppendixCylinder} corresponds to two equations to solve, one arising from $\mu=z$ and the other arising from $\mu=\theta$.   
In steady state, we can evaluate the $z$ equation with the result that
\begin{equation}
\frac{\partial \rho}{\partial z}=0
\end{equation}
implying a constant density. We can then evaluate the $\mu=\theta$ equation, leading to
\begin{equation}
\alpha \left(\rho_0-\rho_{c}\right) -\beta R^{2}\left(v^{\theta}\right)^{2}=0,
\end{equation}
and thus to
\begin{equation}
v^{\theta}_{ss}= \sqrt{{\alpha \left(\rho_0-\rho_{c}\right) \over \beta R^2 }}.
\end{equation}
To find the actual magnitude of the velocity in physical units, we must
contract with the metric tensor using
\begin{equation}
|{\bf v}_{ss}|^2 =g_{\mu \nu} v^{\mu}_{ss}v^{\nu}_{ss}.
\end{equation}
Further, since the only nonzero component of the velocity is $v_{ss}^{\theta}$,
this implies that $|{\bf v}_{ss}|^2 =g_{\theta \theta} v_{ss}^{\theta}v_{ss}^{\theta}$.
Given that $g_{\theta \theta} =R^2$, we can now write the velocity as
\begin{equation}
\left|\mathbf{v}_{s s}\right|^{2}={\alpha \left(\rho_0-\rho_{c}\right) \over \beta }.
\end{equation}
This solution represents a simple circumferential flow around the cylinder
coupled to a uniform density field.  

It is one thing to obtain steady-state solutions.  It is quite another to work out the time evolution
of the solutions.  We now revisit the Toner-Tu dynamics on the cylindrical surface for
the case in which we prescribe a highly-symmetric initial condition with the same symmetry
of circumferential flow and uniform density $\rho_0$, and we work out the relaxation dynamics to the steady state that were shown
in Figure~\ref{fig:DynamicsCylinder}. 
For this geometry, the dynamical equation of interest is given by
\begin{equation}
\frac{\partial v^{\theta}}{\partial t}=\left[\alpha \left(\rho -\rho_{c}\right) -\beta R^{2}\left(v^{\theta}\right)^{2}\right] v^{\theta}-\frac{\sigma}{R} \frac{\partial \rho}{\partial \theta}-\lambda v^{\theta} \frac{\partial v^{\theta}}{\partial \theta}.
\end{equation}
For the special case of high symmetry considered here, we know that both $v^{\theta}$ and $\rho$ do {\it not}
depend upon $\theta$. Thus, the dynamical equation simplifies to the form
\begin{equation}
\frac{d v^{\theta}}{d t}=\left[\alpha \left(\rho-\rho_{c}\right) -\beta R^{2}\left(v^{\theta}\right)^{2}\right] v^{\theta}.
\end{equation}
This nonlinear equation is separable resulting in the elementary integral
\begin{equation}
{1 \over \beta R^2} {dv^{\theta} \over v^{\theta} \left({\alpha \left(\rho-\rho_{c}\right) \over \beta R^2} - (v^{\theta})^2    \right)}
=dt.
\end{equation}
Given an initial magnitude $v(0)$ = $R~ v^{\theta}(0)$ and a uniform density $\rho_0$, the solution for the time dependent relaxation is given
by 
\begin{equation}
\left|\mathbf{v}\right|^{2} = R^2 (v^{\theta})^{2}=\frac{\frac{\alpha \left(\rho_0-\rho_{c}\right)} {\beta }}{1-\left(1-\frac{\alpha \left(\rho_0-\rho_{c}\right)}{\beta ~ v(0)^{2}}\right) e^{-2 \alpha \left(\rho_0-\rho_{c}\right) t}}.
\end{equation}
In Figure~\ref{fig:DynamicsCylinder}, we compared the dynamics predicted here analytically to our finite element method solution of the same problem.

\subsection{Analytical Solution for Toner-Tu on the Sphere}
\label{SectionAppendixSphereAnalytical}

We next take up the analysis of the minimal Toner-Tu model on the sphere.
Previous work \cite{Sknepnek2015, Shankar2017} described a highly-symmetric rotating band solution for flocks on the sphere, in which both the density and velocity depend only upon the azimuthal angle $\theta$ at steady-state. Here, we present the analytical solution for density and velocity on the sphere derived by Shankar {\it et al.} \cite{Shankar2017}, which is compared to our numerical results in Section \ref{Section:AnalyticSolutions}. For their steady-state analysis, Shankar {\it et al.} \cite{Shankar2017} make the approximation that the neighbor coupling term with coefficient $D$ is absent, resulting
in the simpler minimal version of the steady-state Toner-Tu equations of the form
\begin{equation}
\lambda v^{\nu} \nabla_{\nu} v^{\mu}=\left[\alpha \left(\rho-\rho_{c}\right)- \beta g_{ab} v^{a} v^{b}\right] v^{\mu}-\sigma \nabla^{\mu} \rho.
\label{eqn:TonerTuSphereSSAppendix}
\end{equation}
We note that while Shankar {\it et al.} chose the field variables $\rho$ and ${\bf p}=\rho {\bf v}$, we continue in the language of  $\rho$ and ${\bf v}$.
To concretely solve these equations, the  first step is the parameterization of the spherical surface resulting in
\begin{equation}
{\bf r}(\theta, \phi)=(R \cos \phi \sin \theta, R \sin \phi \sin \theta, R \cos \theta).
\end{equation}
Next we need to find the tangent vectors which are defined
via
\begin{equation}
{\bf a}_i = \frac{\partial {\bf r}}{\partial u^i},
\end{equation}
where $u^i$ is the ``coordinate'' of the surface defined above.
Given these definitions, we have 
\begin{equation}
{\bf a}_{\theta}=(R \cos \phi \cos \theta, R \sin \phi \cos \theta,-R \sin \theta)
\end{equation}
and
\begin{equation}
{\bf a}_{\phi}={\partial {\bf r} \over \partial \phi} =(-R \sin \phi \sin \theta, R \cos \phi \sin \theta, 0)
\end{equation}
as our two tangent vectors, exactly as expected.   
The components of the  metric tensor are obtained by computing the dot product of these
tangent  vectors and are given as
\begin{equation}
{\bf g}=
\left[\begin{array}{ll}
R^{2} & 0 \\
0 & R^{2} \sin ^{2} \theta
\end{array}\right].
\end{equation}
with much of the simplicity resulting from the fact that the basis vectors are orthogonal.
In moving back and forth between covariant and contravariant components of our
vectors of interest, we also need 
\begin{equation}
g^{ab}=\left(g_{a b}\right)^{-1}=\left[\begin{array}{cc}
{1 \over  R^{2}} & 0 \\
0 & \frac{1}{R^{2} \sin ^{2} \theta}
\end{array}\right].
\end{equation}

Eqn.~\ref{eqn:TonerTuSphereSSAppendix} corresponds to  two equations we have to solve, one arising from $\mu=\theta$ and the other arising from $\mu=\phi$.   
To evaluate the covariant derivatives on the surface present in these equations, we will need the Christoffel symbols.
Generically, the Christoffel symbols are defined intuitively to tell us how our basis vectors change
when we take small excursions with the parameters that characterize our
surface of interest.  Specifically, we have
\begin{equation}
\Gamma_{ij}^k={\partial {\bf a}_i \over \partial u^j} \cdot {\bf a}^k,
\end{equation}
measuring how an excursion in $u^j$ leads to a change of ${\bf a}_i$.
For the highly symmetric geometry of the cylinder considered earlier, the Christoffel symbols all
vanish.  We can see this by recognizing that the z-direction and the $\theta$ direction
are essentially uncoupled.  That is, there is no dependence of ${\bf a}_{\theta}$ on
the coordinate $z$ and similarly, there is no dependence of ${\bf a}_z$ on $\theta$.
Further, $\partial {\bf a}_{\theta}/\partial \theta$ is perpendicular to ${\bf a}^{\theta}$
and hence there is no contribution to the Christoffel symbol.   

Evaluation of
the covariant derivatives for the sphere requires us to invoke  Christoffel symbols which
unlike the cylinder case, do not all vanish.  Because of the symmetry of the sphere, only a few of the Christoffel symbols survive.  Using the tangent vectors defined
above, we have
\begin{equation}
\Gamma_{\phi \phi}^{\theta}={\partial {\bf a}_{\phi} \over \partial \phi} \cdot {\bf a}^{\theta}= -\sin \theta \cos \theta
\end{equation}
and
\begin{equation}
\Gamma_{\theta \phi}^{\phi}={\partial {\bf a}_{\theta} \over \partial \phi} \cdot {\bf a}^{\phi}=\cot \theta.
\end{equation}
With these geometrical preliminaries settled, we can now turn to the steady-state form of the dynamical equations themselves.

For the case when $\mu=\theta$, the Toner-Tu equations become
\begin{equation}
\lambda v^{\nu} \nabla_{\nu} v^{\theta}=-\sigma \nabla^{\theta} \rho
\end{equation}
since the term 
$\left[\alpha\left(\rho-\rho_{c}\right)-\beta g_{a b}  v^{a} v^{b}\right] v^{\theta}$ vanishes because $v^{\theta}=0$
as a result of our assumption that the solution only has a $\phi$ component of velocity. 
At first, we might be tempted to set the term $\lambda v^{\nu} \nabla_{\nu} v^{\theta}$ equal to zero,
but because of the covariant derivative, there is a contribution due to $v_{ss}^{\phi}$.
To see that, we note that the full term $\lambda v^{\nu} \nabla_{\nu} v^{\theta}$ requires a covariant derivative and is given by
\begin{equation}
\lambda v^{\nu} \nabla_{\nu} v^{\theta}=\lambda v^{\nu} \partial_{\nu} v^{\theta}+\lambda v^{\nu} \Gamma_{\alpha \nu}^{\theta} v^{\alpha}.
\end{equation}
The first term on the right vanishes because $v^{\theta}=0$.
The term $\lambda v^{\nu} \Gamma_{\alpha \nu}^{\theta} v^{\alpha}$ results in
$\lambda v^{\phi} \Gamma_{\phi \phi}^{\theta} v^{\phi} =- \lambda \sin \theta \cos \theta\left(v_{s s}^{\phi}\right)^{2}$.
A critical point necessary for evaluating
$-\sigma \nabla^{\theta} \rho$ is the relation between $\nabla_{\mu}$ and $\nabla^{\mu}$.  Specifically, we note
that 
\begin{equation}
\nabla^{\mu}= g^{\mu \nu} \nabla_{\nu}.
\end{equation}
In light of this relationship, we have
\begin{equation}
\nabla^{\theta}=g^{\theta j} \nabla_{j}= g^{\theta \theta} \nabla_{\theta}.
\end{equation}
Given that $g^{\theta \theta} =1/R^2$ and $\nabla_{\theta}=\partial/\partial \theta$, we can now write out
the steady-state $\mu=\theta$ equation as
\begin{equation}
\lambda \sin \theta \cos \theta\left(v_{s s}^{\phi}\right)^{2}=\frac{\sigma}{R^{2}}{\partial \rho_{ss} \over \partial \theta}.
\label{eqn:SteadyStateMuTheta}
\end{equation}
For the case when $\mu=\phi$, we have
\begin{equation}
\lambda v^{\nu} \nabla_{\nu} v^{\phi}=\left[\alpha \left(\rho-\rho_{c}\right)-\beta g_{a b}  v^{a} v^{b}\right] v^{\phi},
\label{eqn:SteadyState1}
\end{equation}
where we have dropped the  term $\sigma \nabla^{\phi} \rho$ term since by symmetry, we have {\it assumed}
that the density does not depend upon the coordinate $\phi$.  The first term vanishes, resulting in
\begin{equation}
v_{s s}^{\phi}\left[\alpha\left(\rho_{s s}-\rho_{c}\right)-\beta R^{2} \sin ^{2} \theta\left(v_{s s}^{\phi}\right)^{2}\right]=0.
\label{eqn:SteadyState2}
\end{equation}
We can now use these two conditions in our two unknowns $\rho_{ss}(\theta)$ and $v_{ss}^{\phi}(\theta)$ to solve for
the steady-state values of density and velocity.  Shankar et al. \cite{Shankar2017} begin by defining a  density $X(\theta)$ with a new reference value as 
\begin{equation}
X(\theta)=\rho_{s s}(\theta)-\rho_{c}.
\end{equation}
We can use this definition to solve eqn.~\ref{eqn:SteadyState2} for $v_{ss}^{\phi}$ as  
\begin{equation}
(v_{ss}^{\phi})^2 = {\alpha X \over \beta R^2 \sin ^2 \theta}.
\end{equation}
This can now be substituted into eqn.~\ref{eqn:SteadyStateMuTheta} to obtain the differential equation
\begin{equation}
{\sigma \over R^2} {dX \over d \theta} = {\lambda \alpha \over \beta R^2} {\mbox{cos} ~\theta \over \mbox{sin} ~\theta} X.
\end{equation}
Multiplying both sides by $R^2/\sigma$ results in the emergence of a 
dimensionless parameter 
given by
\begin{equation}
\eta=\frac{\lambda \alpha}{\beta \sigma}
\end{equation}
permitting us to rewrite the equation for $X(\theta)$ as 
\begin{equation}
{dX \over d\theta} = \left( \eta~ \mbox{cot}~\theta \right) X.
\end{equation}
This differential equation is separable 
and upon integration yields the simple expression
\begin{equation}
\mbox{ln}~X = \eta~\mbox{ln}~(\mbox{sin}~\theta)+C.
\end{equation}
By exponentiating both sides of the equation, we then find the solution
\begin{equation}
X(\theta)= A~\mbox{sin}^{\eta}\theta.
\end{equation}
The unknown coefficient $A$ can be determined by exploiting the symmetry of the solution
that tells us that the maximal density will be at the equator of the sphere, implying
\begin{equation} 
X({\pi \over 2}) = \rho_{ss}({\pi \over 2}) - \rho_c = A~\mbox{sin}~{\pi \over 2},
\end{equation}
permitting us then to write the 
steady-state solutions for the density as
\begin{equation}
\rho_{s s}(\theta)=\rho_{c}+\left(\rho_{0}-\rho_{c}\right) A_{\eta} \sin ^{\eta} \theta
\end{equation}
where the prefactor $A_{\eta}$ is defined by
\begin{equation}
A_{\eta}=2 \Gamma((3+\eta) / 2) /[\sqrt{\pi} \Gamma(1+\eta / 2)].
\end{equation}
Note that the dimensionless parameter $\eta$ gives us a convenient knob to tune in our calculations, which allowed us to directly compare with the numerical results of our finite element calculations
in Figure~\ref{fig:rhossCOMSOLvsAnalytical}.  

We can also work out the steady-state value of the velocity field by invoking
the condition
\begin{equation}
(v_{ss}^{\phi})^2 ={\alpha \over \beta R^2 \mbox{sin}^2\theta} X(\theta).
\end{equation}
Recall that  to find the actual magnitude of the velocity in physical units, we must
contract with the metric tensor using
\begin{equation}
|{\bf v}_{ss}|^2 =g_{\mu \nu} v^{\mu}v^{\nu}.
\end{equation}
Further, since the only nonzero component of the velocity is $v_{ss}^{\phi}$,
this implies that $|{\bf v}_{ss}|^2 =g_{\phi \phi} v_{ss}^{\phi}v_{ss}^{\phi}$
Given that $g_{\phi \phi} =R^2 \mbox{sin}^2~\theta$, we can now write the velocity as
\begin{equation}
\left|\mathbf{v}_{s s}\right|^{2}=\frac{\alpha}{\beta}\left(\rho_{0}-\rho_{c}\right) A_{\eta} \sin ^{\eta} \theta.
\end{equation}
For the specific case of $\eta=2$, for example, we find
\begin{equation}
A_2={2 \Gamma({5 \over 2}) \over \sqrt{\pi} \Gamma(2)} = {3 \over 2}.
\end{equation}
Using this value for $A_2$, we can then write the density and the velocity fields as
\begin{equation}
\rho_{ss}(\theta)=\rho_c+{3 \over 2} (\rho_0-\rho_c) \mbox{sin}^2\theta
\end{equation}
and
\begin{equation}
\left|\mathbf{v}_{s s}\right|^{2}={3 \over 2} (\rho_0-\rho_c) \mbox{sin}^2\theta,
\end{equation}
respectively. 
These solutions from the work of Shankar {\it et al.} \cite{Shankar2017} provided an opportunity to test our numerical
implementation of the Toner-Tu theory on a curved surface using the finite element method, as shown in Figure~\ref{fig:rhossCOMSOLvsAnalytical} of Section \ref{SectionSphereAnalytical}.

\bigskip
\newpage
\noindent{\bf Video Legends}\\
\addcontentsline{toc}{section}{Video Legends}

{\small 
	\textbf{Video 1. Toner-Tu herding in channels with obstacles.} Starting from uniform density and a disordered velocity field, density (colors) and velocity (black arrows) evolve according to the minimal Toner-Tu equations in 2D channels with periodic boundary conditions at their ends. The presence of an obstacle in the channel can lead to unidirectional flow around the obstacle (middle) or direction reversals due to herd reflection off the obstacle (bottom), depending on the obstacle size.
	
	\medskip
	\textbf{Video 2. Toner-Tu herd dynamics on the sphere.} Dynamics of herd density (colors) and velocity (black arrows) on the sphere, during the evolution from an initial condition of uniform density and circumferential velocity $v_{\phi}= v_{\mbox{pref}} \sin \theta$ to the steady-state band pattern that emerges in the long-time limit. The steady-state band emerges through the damping of density oscillations by the neighbor coupling term $D \nabla^2 {\bf v}$.

	\medskip
	\textbf{Video 3. Herds on the sphere can form rotating patch, oscillating band, and steady-state band patterns.} In these examples, herd dynamics are initialized by confining wildebeest to an angular wedge between $\phi=0$ and $\phi=\pi/2$ and giving them a circumferential velocity $v_{\phi}= v_{\mbox{pref}} \sin\theta$. With a pressure term coefficient of $\sigma = 5 ~{\mbox m^4/s^2}$, a rotating patch pattern of herd density emerge (top). A larger pressure term ($ \sigma = 15 ~{\mbox m^4/s^2}$) leads the density field to exhibit temporal oscillations about the equatorial plane (middle). Given an ever larger pressure term  ($ \sigma = 20 ~{\mbox m^4/s^2}$), herd density is restricted to a narrow range around $\rho_{0} ~(0.0625 ~ {\mbox m^{-2}})$, and a steady-state band pattern of density emerges. This video shows the dynamics in the long-time limit, for a snapshot from time t = 100,000 s - 105,000 s.

	\medskip
	\textbf{Video 4. Dynamics of a Toner-Tu herd on an undulating island landscape}. The simulation is initialized with uniform density $\rho_{0} = 0.25~{\mbox m^{-2}}$ and a disordered velocity field. Comparing the top simulation ($\zeta = 0~ {\mbox m/s^2}$) and the bottom ($\zeta = 0.05~ {\mbox m/s^2}$) highlights the contribution of the gravitational force. In the presence of the gravitational force, the hills act as soft obstacles, favoring herd circulation around them.

}

\bigskip


\bigskip
\bigskip

\noindent{\bf Acknowledgements}\\
\addcontentsline{toc}{section}{Acknowledgements}

\noindent We are deeply grateful to a number of generous colleagues who have
carefully described their thinking and work on the formulation of
continuum descriptions on surfaces, and to others who have provided invaluable feedback on this work.  We especially thank Ashutosh Agarawal, Marino Arroyo, David Bensimon, Mark Bowick, Markus Deserno, Soichi Hirokawa, Greg Huber, Alexei Kitaev (who showed us how to solve the half space problem), Jane Kondev, Elgin Korkmazhan, Deepak Krishnamurthy, Kranthi Mandadapu, Madhav Mani, Cristina Marchetti, Walker Melton, Alexander Mietke, Phil Nelson, Silas Nissen, Manu Prakash, Sriram Ramaswamy, Suraj Shankar, Mike Shelley, 
Sho Takatori,  John Toner, Yuhai Tu, and Vincenzo Vitelli. We thank Nigel Orme for his work on Figures 1-3. This work was supported by a Burroughs Wellcome Career Award at the Scientific Interface (C.L.H.), NIH R35GM130332 (A.R.D.), an HHMI Faculty Scholar Award (A.R.D.), NIH MIRA 1R35GM118043 (R.P.), and the Chan Zuckerberg Biohub (R.P.). C.L.H. is a Damon Runyon Fellow supported by the Damon Runyon Cancer Research Foundation (DRG-2375-19).”\\

\bigskip
\bigskip

\noindent{\bf Competing interests:}\\

\noindent Authors declare that they have no competing interests.

\bigskip

\newpage
\addcontentsline{toc}{section}{References}

\bibliographystyle{model1-num-names}

\end{document}